\documentclass[10pt, journal, letterpaper]{IEEEtran} 
\usepackage{amsmath,amsfonts}
\usepackage{algorithm}
\usepackage{array}
\usepackage[caption=false,font=normalsize,labelfont=sf,textfont=sf]{subfig}
\usepackage{textcomp}
\usepackage{stfloats}
\usepackage{verbatim}
\usepackage{url}
\usepackage{verbatim}
\usepackage{graphicx}
\usepackage{cite}
\usepackage{booktabs}  % 
\usepackage{algorithm}
\usepackage{algpseudocode}  % 用于 algorithmic 环境
\usepackage{amsmath}        % 用于 math 符号，如 \mathsf, \boldsymbol 等
\usepackage{amssymb}        % 支持更多数学符号
\usepackage{subfig}
\usepackage{graphicx}       % 导入graphicx宏包
\usepackage{optidef}
\usepackage{multirow}
\usepackage[group-separator={}]{siunitx}
\usepackage{enumitem}
\IEEEoverridecommandlockouts  % 新增：覆盖锁，防止标题空白

\bstctlcite{IEEEexample:BSTcontrol}

\graphicspath{{Fig/Results_Revision_Final/CDF/}，{Fig/Results_Revision_Final/complexity/},{Fig/Results_Revision_Final/loss/},{Fig/SystemModel/}}

\begin{document}
\title{Scalable GNN-Based Power Allocation for Rate-Splitting Cell-Free Massive MIMO Systems}

% \thanks{This paper was produced by }% <-this % stops a space
\author{
Ruomeng~Wang,
Yin~Xu,~\IEEEmembership{Senior~Member,~IEEE},
Aimin~Tang,~\IEEEmembership{Senior~Member,~IEEE},
Xiaowu~Ou,
Jun~Zhu,
Dazhi~He,~\IEEEmembership{Senior~Member,~IEEE},
Lifeng~Wang,~\IEEEmembership{Member,~IEEE},
and Wenjun~Zhang,~\IEEEmembership{Fellow,~IEEE}

% \thanks{
% This work was supported in part by the National Key Research and Development Project of China under Grant 2023YFF0904603; in part by the National Natural Science Foundation of China Program under Grant 62422111. \textit{(Corresponding author: Yin Xu.)}

% C. Zhang, S. Peng, X. Ou, X. Guo, Y. Xu, D. He, and W. Zhang are with the Cooperative Medianet Innovation Center, Shanghai Jiao Tong University, Shanghai 200240, China. Dazhi He is also affiliated with Pengcheng Laboratory, Shenzhen 518055, China (e-mail: \{cixiaozhang, sjtu2019psz, xiaowu\_ou, guoxinghao, xuyin, hedazhi, zhangwenjun\}@sjtu.edu.cn). 

% Y. Zhu is with the National Mobile Communications Research Laboratory, Southeast University, Nanjing 210096, China. And she is also with Purple Mountain Laboratories, Nanjing 211111, China. (email: yongxu.zhu@seu.edu.cn)

% H. Hong is with the Department of Electronic and Electrical Engineering, University College London, Torrington Place, WC1E7JE, United Kingdom (e-mail: hanjiang.hong@ucl.ac.uk). }

\thanks{
(Corresponding author: Yin Xu.)

Ruomeng Wang, Yin Xu, Aimin Tang, Xiaowu Ou, Jun Zhu, Dazhi He, and Wenjun Zhang are with Shanghai Jiao Tong University, Shanghai 200240, China (e-mail: \{wangruomeng, xuyin, tangaiming, xiaowu\_ou, zhujun\_22, hedazhi, zhangwenjun\}@sjtu.edu.cn).

Lifeng Wang is with the School of Future Information Technology, Fudan University, Shanghai 200433, China (e-mail: lifengwang@fudan.edu.cn).
}}

% \markboth{}%

% Remember, if you use this, you must call \IEEEpubidadjcol in the second

\maketitle
\begin{abstract} Cell-free massive multiple-input multiple-output (CF-mMIMO) systems provide enhanced coverage and capacity for next-generation wireless networks. However, CF-mMIMO systems face significant challenges in downlink power allocation (PA) due to imperfect channel state information (CSI), severe multi-user interference (MUI), and high computational complexity. To address these issues, rate-splitting multiple access (RSMA) is adopted as a robust interference management strategy. Accordingly, this paper proposes an unsupervised and scalable graph neural network (GNN) framework for PA in rate-splitting CF-mMIMO (RS-CF-mMIMO) systems, relying exclusively on large-scale fading (LSF) coefficients without instantaneous CSI. To resolve the dimensionality mismatch in dynamic networks, we introduce a slice-based adaptive layer that projects variable-dimension features into a fixed latent space. This mechanism enables a unified model to generalize across diverse topologies without retraining. Within this architecture, the sum spectral efficiency (SE) is maximized under per-AP power constraints, assuming maximum-ratio precoding for common streams and regularized zero-forcing precoding for private streams. We also derive a weighted minimum mean-square error-alternating direction method of multipliers (WMMSE-ADMM) algorithm as a performance upper bound. Extensive simulations verify that the proposed GNN framework achieves near-optimal SE and outperforms unsupervised deep neural networks (DNNs) across diverse system sizes and pilot assignment schemes. Furthermore, the scalable variant maintains robust performance while reducing the trainable parameter count by over 57\% relative to DNNs and decreasing inference latency by up to three orders of magnitude compared with WMMSE-ADMM.
\end{abstract}


\section{Introduction}
\IEEEPARstart{M}{assive} multiple-input multiple-output (mMIMO) is an essential technology designed for facilitating high-density user equipments (UEs) in next-generation wireless networks, delivering substantial gains in both spectral efficiency (SE) and energy efficiency (EE)~\cite{2014mMIMO,2024mMIMO}. However, its performance within conventional cellular networks is limited by inherent architecture drawbacks, including uneven coverage and severe inter-cell interference. To address these issues, cell-free massive MIMO (CF-mMIMO) is proposed by distributing access points (APs) within a specific coverage area to collaboratively serve multiple UEs without creating artificial cell boundaries~\cite{2021JCIN_He}. This paradigm exploits the spatial diversity and multi-user interference (MUI) suppression capabilities of the system, thereby providing near-uniform quality of service (QoS) to all UEs~\cite{2024CFmMIMO,2024CFmMIMO2,2024CFmMIMO3,2024CFmMIMO4}. Nevertheless, practical CF-mMIMO systems also face several challenges that must be resolved to fully realize their potential, primarily the high computational complexity of signal processing at central processing units (CPUs) and complex interference management issues caused by pilot contamination~\cite{2022CST_Ammar, 2020TCOM_Bjorson, 2020TWC_Bjornson,2022TWCGarg}.

In particular, pilot contamination gains
significant attention in recent years, as it induces channel
estimation errors that severely degrade SE performance, further exacerbated in high-density user scenarios, such as massive machine-type communications (mMTC), where pilot resources are severely constrained~\cite{2022TWC_Ancup}. 
To mitigate these impairments, rate-splitting multiple access (RSMA) has emerged as a powerful paradigm for robust interference management. Unlike conventional space-division multiple access (SDMA) or non-orthogonal multiple access (NOMA), RSMA provides a generalized transmission framework that bridges the gap between fully decoding interference and treating it strictly as noise, offering unprecedented flexibility and robustness against the imperfect channel state information (CSI) prevalent in large-scale deployments~\cite{2022RSMAsurvey,2023RSMAsurvey}. Specifically, in rate-splitting CF-mMIMO (RS-CF-mMIMO) systems, the message of each UE is split into common and private parts. The former are aggregated into a shared common stream, while the latter are mapped to dedicated private streams for individual decoding. Through the optimization of power allocation (PA) for split messages, this framework facilitates partial decoding of interference while treating the remaining interference as noise~\cite{2018JWCN_Ymao}. Therefore, PA is pivotal in RS-CF-mMIMO systems, dictating both the interference mitigation efficacy and the overall SE~\cite{2024TWCGottsch,2022Sustainability_Imoize}. 

Prior research extensively investigates conventional optimization techniques to tackle the PA problem and characterize fundamental performance limits~\cite{2017TWC_Ngo, 2024IoT_Chongzheng_H, 2017TWC_Elina, 2025TWCWang, 2023WCL_Facheng_L, 2023Tcom_Flores}. For instance, the work in~\cite{2017TWC_Ngo} introduces a max-min fairness PA algorithm by solving a sequence of second-order cone programs (SOCP) for the downlink, albeit at the cost of the computational complexity that scales polynomially with the system size. To accelerate convergence, some works compress the input data at the cost of information loss. In~\cite{2024IoT_Chongzheng_H}, the authors propose an accelerated projected gradient method for large-scale CF-mMIMO systems, achieving SE performance comparable to successive convex approximation (SCA). Alternatively, other studies simplify the PA optimization by introducing tractable assumptions. For example, the authors in~\cite{2017TWC_Elina} simplify the signal-to-interference-plus-noise ratio (SINR) expression under the assumptions of conjugate beamforming and full-power transmission. Additionally, several approaches decompose the original problem into subproblems with closed-form solutions. For instance, a randomized alternating direction method of multipliers (ADMM) algorithm is developed in~\cite{2025TWCWang}, which requires solving only a small number of subproblems per iteration. Similarly, a closed-form precoder for the common and private streams for RS-CF-mMIMO systems is derived under fixed system configurations~\cite{2023WCL_Facheng_L,2023Tcom_Flores}.

To further alleviate the computational burden, recent works in~\cite{2022Globalcomm_Salaun,2023TWC_Bjornson,2021TWC_HLee,2021ICC_NRajapaksha,2024OJCS_Muhammad,2023WCL_NRajapaksha,2022ComL_HH,2025Tcom_DY,mousavi2025federated, irkicatal2024deep,johnston2024rnn,2021Globalcomm_Yang_Z,2023TVT_Yongshun_Z,2020Access_Yu_Z} explore deep learning-based approaches to approximate near-optimal power coefficients. Under supervised frameworks, a convolutional neural network (CNN) is proposed in~\cite{2022Globalcomm_Salaun} to learn max-min PA solutions with labels generated by SOCP or SCA solvers, whereas distributed deep neural networks (DNNs) at APs are employed in~\cite{2023TWC_Bjornson} to approximate weighted minimum mean square error (WMMSE) strategies. Similarly, a supervised framework is developed in~\cite{2021TWC_HLee} to jointly optimize fronthaul interactions and decentralized edge computation. Conversely, unsupervised approaches are also investigated in~\cite{2021ICC_NRajapaksha,2024OJCS_Muhammad,2023WCL_NRajapaksha}, which directly optimize DNNs for max-min or sum-rate objectives, jointly tackling user association and pilot assignment. To further reduce signaling overhead within this unsupervised paradigm, a decentralized beamforming scheme is proposed in~\cite{2022ComL_HH}, while~\cite{2025Tcom_DY} introduces a model-based unrolled WMMSE algorithm to balance fronthaul constraints and sum-rate performance. Extending these paradigms to RSMA-specific resource management, recent research further employs federated DRL~\cite{mousavi2025federated}, multi-agent DRL~\cite{irkicatal2024deep}, and recurrent neural networks (RNNs)~\cite{johnston2024rnn} to jointly optimize user selection, precoding, and PA coefficients. To address the scalability bottleneck inherent in these approaches, some researchers introduce techniques such as aggregating large-scale fading (LSF) coefficients~\cite{2021Globalcomm_Yang_Z,2023TVT_Yongshun_Z} or substituting fully connected layers with convolutional layers~\cite {2020Access_Yu_Z}.

Beyond DNN-based schemes, graph neural networks (GNNs) are leveraged to capture topological dependencies and interactions between APs and UEs, offering superior data efficiency and generalization capabilities. Specifically, prior studies indicate that GNNs with $n$ nodes can achieve $O(n)$ times lower generalization error and require $O(n^2)$ times fewer training samples than multilayer perceptrons (MLPs)~\cite{2022arXiv_GraphRepresentation,2020arXiv_GraphNNforScalable}. Capitalizing on such graph-native features, recent works develop GNN-based strategies for the PA task~\cite{2024ICMLCN,2024TVT,2022Tcom_Ngo,2024arXiv_Benjamin, huang2025gnn}. For instance, SINRnet is proposed to maximize EE, where links are modeled as nodes and channels are encoded into four types via edge attributes based on their roles in the SINR expression~\cite{2024ICMLCN}. Parallel work in~\cite{2024TVT} develops a distributed GNN to maximize the sum ergodic SE relying on local CSI and limited information exchange among APs. Further contributions include geometric-mean rate maximization for balanced user rates~\cite{2022Tcom_Ngo} and supervised GNNs for optimal linear precoding prediction~\cite{2024arXiv_Benjamin}. Regarding RSMA-specific applications, a GNN-aided framework is introduced in hardware-impaired cell-free ultra-reliable low-latency communication (URLLC) systems~\cite{huang2025gnn}, where nodes represent AP-UE associations and edges characterize MUI.

Recent studies explore various strategies to enable GNN scalability across dynamic system configurations~\cite{2024TWC_Shashwat, sun2024resource, 2021TVT_Yu_Z, 2024TWC_Bohan}. Specifically, some works employ one-hot-encoded edge attributes to represent per-AP serving densities~\cite{2024TWC_Shashwat}, while others utilize attention mechanisms to dynamically weight neighbor contributions~\cite{sun2024resource}. However, these approaches lead to a significant increase in the number of trainable parameters as the network scales, rendering them inefficient for dense deployments. Alternatively, scalability is attempted by compressing the graph structure, either through dynamic pooling mechanisms that aggregate nodes, or by selecting the top-$K$ ``closest'' AP-UE pairs to form a fixed-size edge set~\cite{2024TWC_Bohan}. Although these achieve system-size independence, they introduce additional preprocessing overhead and generalize poorly to unseen system configurations.

In light of the aforementioned challenges, we propose a GNN-based PA framework designed to be robust against interference, efficient in training, and inherently scalable to dynamic system sizes. To mitigate the severe pilot contamination and MUI arising from imperfect CSI, we incorporate RSMA as a robust interference management strategy via common-private message splitting. Distinct from conventional supervised approaches, we formulate an SE-oriented objective function to drive end-to-end unsupervised learning, thereby eliminating the computational burden of label generation. 

Notably, to achieve size-agnostic scalability, we devise a novel slice-based adaptive embedding mechanism that addresses two critical limitations of conventional zero-padding methods. Firstly, in terms of computational efficiency, zero-padding couples the model complexity to the maximum predefined network size, leading to parameter redundancy and unnecessary computations~\cite{Mit2026zeropad}. In contrast, our mechanism activates only the weights associated with the actual AP and UE dimensions of each network realization, so that the computational cost scales with the active topology. Secondly, in terms of topological fidelity, padded zeros may create artificial ``ghost nodes'' that disrupt the graph Laplacian's symmetric normalization~\cite{Liu2022zeropad}, while our slice-based mechanism dynamically maps variable-dimensional features into a fixed latent space, completely obviating ghost nodes and empowering a single unified model to generalize seamlessly across diverse dynamic topologies without retraining.

The main contributions of this paper are summarized as follows:

\begin{itemize}
    \item[$\bullet$] We integrate RSMA into the CF-mMIMO system to ensure robust interference management and formulate the downlink PA for common and private streams as an unsupervised graph learning task. By modeling APs and UEs as heterogeneous nodes and large-scale links as the edges, the proposed framework relies exclusively on scaled LSF coefficients, thereby eliminating the dependency on instantaneous CSI while maximizing system SE under per-AP power constraints via an end-to-end objective.
    
    \item[$\bullet$] We propose a scalable GNN architecture featuring a novel slice-based adaptive embedding layer. This mechanism implements a topology-aware projection by dynamically extracting a sub-matrix from a global weight pool to map variable inputs to a fixed latent space. Crucially, this design obviates the need for zero-padding and effectively decouples the model complexity from the worst-case system size, empowering a single unified model to generalize across heterogeneous configurations without retraining.

    \item[$\bullet$] We derive a WMMSE reformulation for the considered problem and develop an ADMM-based solver to serve as a performance upper bound. Extensive simulations are conducted to assess both SE and computational complexity across diverse system sizes and pilot assignment schemes, demonstrating near-optimal performance with substantially reduced trainable parameter count and millisecond-level inference latency.
\end{itemize}

\subsubsection{Organizations}
The remainder of the paper is organized as follows: the system model is described in Section II, followed by the problem formulation and the derivation of the WMMSE–ADMM algorithm in Section III. The proposed scalable GNN framework is detailed in Section IV. Numerical results and a complexity analysis are presented in Section V, and the paper concludes in Section VI.

\subsubsection{Notations}
Bold uppercase and bold lowercase letters denote matrices and vectors, respectively. The spaces $\mathbb{C}^{m\times n}$ and $\mathbb{R}^{m\times n}$ represent the sets of $m\times n$ complex-valued and real-valued matrices. $\mathbf{I}_M$ and $\mathbf{0}_{m\times n}$ denote the $M\times M$ identity matrix and the $m\times n$ zero matrix. The superscripts $(\cdot)^{{T}}$, $(\cdot)^{{H}}$, and $(\cdot)^{-1}$ stand for the transpose, Hermitian transpose, and matrix inverse operations. Regarding mathematical operators, $\mathbb{E}[\cdot]$, $\Re(\cdot)$, $\mathrm{tr}(\cdot)$, $\|\cdot\|_2$, $|\cdot|$, and $\circ$ denote the expectation, real part, trace, Euclidean norm, cardinality, and Hadamard product, respectively. Finally, $\mathcal{CN}(\mu, \sigma^2)$ denotes the circularly symmetric complex Gaussian distribution with mean $\mu$ and variance $\sigma^2$.

\section{System and Signal Models}\label{S_system}
\subsection{System Model}
\begin{figure}[t!]
\centering
% trim=0 0 2cm 0 表示切掉右边 2cm 的空白，clip 必须加上，否则只移动不裁剪
\includegraphics[width=3in, trim=0 0 0.85cm 0, clip]{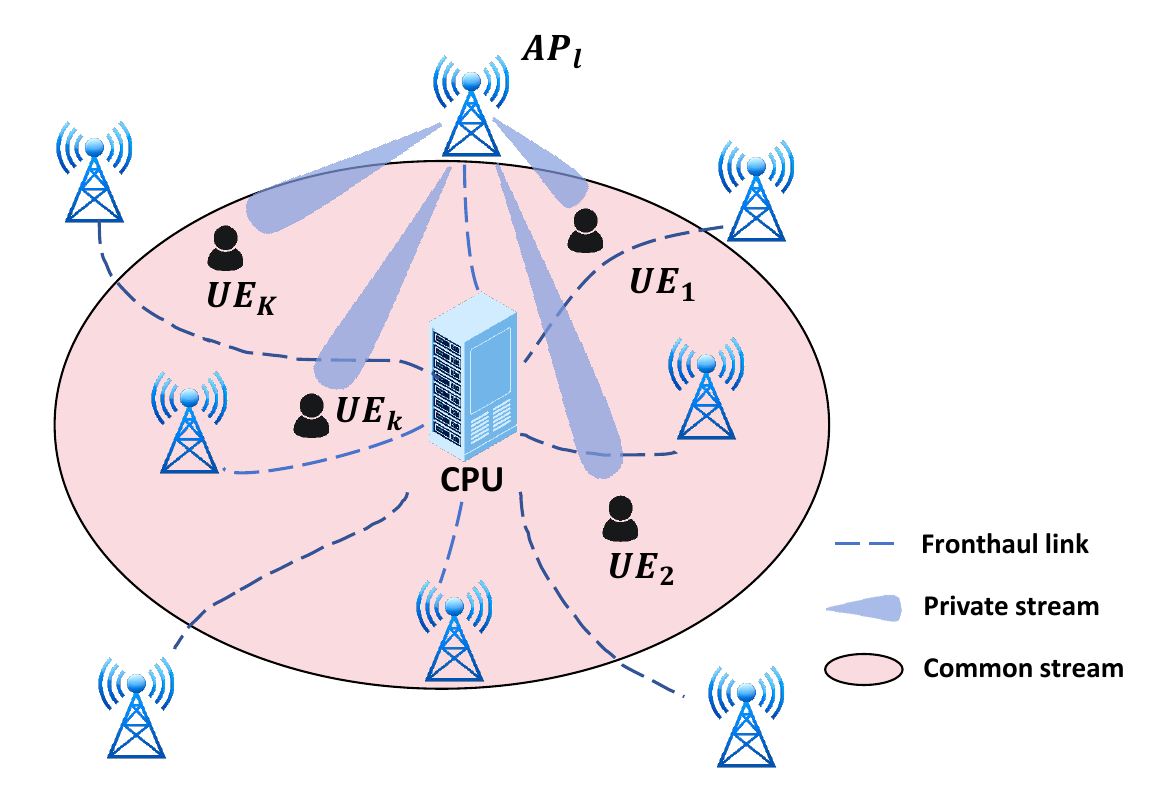}
\caption{The CF-mMIMO system architecture.}
\label{fig_1}
\end{figure}

As illustrated in Fig.~\ref{fig_1}, this paper considers a CF-mMIMO system. It consists of a central processing unit (CPU) deployed at the edge server, and $L$ distributed APs. Each AP is equipped with $N$ antennas and connected to the CPU via fronthaul links to serve $K$ single-antenna UEs coherently. All centralized operations, including precoding and power allocation, are performed at the CPU and then communicated to the respective APs for coherent transmission.

The channel vector between AP~$l$ and UE~$k$ is denoted by ${\mathbf{h}}_{k,l} \in \mathbb{C}^{N \times 1}$ and is modeled as correlated Rayleigh fading
\begin{equation}
    {\mathbf h}_{k,l} \sim \mathcal{CN}({\mathbf 0},{\mathbf R}_{k,l}),
\end{equation}
where ${\mathbf{R}}_{k,l} \in {\mathbb{C}}^{N \times N}$ is the spatial correlation matrix, and the LSF coefficient is given by $\beta_{k,l}=\frac{1}{N}\operatorname{tr}({\mathbf{R}}_{k,l})$.

The system operates under a time division duplex (TDD) protocol, leveraging channel reciprocity to estimate downlink channels from uplink pilots. We adopt a standard block fading channel model where the channel remains static and frequency-flat within a coherence block of length $\tau_\mathrm{c}$ symbols. Each block is partitioned into $\tau_\mathrm{p}$ symbols for uplink pilot transmission and $\tau_\mathrm{d}$ symbols for downlink data transmission, satisfying $\tau_\mathrm{c} = \tau_\mathrm{p} + \tau_\mathrm{d}$. Regarding the uplink pilot training, let $\boldsymbol{\Phi} = [\boldsymbol{\phi}_1, \dots, \boldsymbol{\phi}_{\tau_\mathrm{p}}] \in \mathbb{C}^{\tau_\mathrm{p} \times \tau_\mathrm{p}}$ denote the set of $\tau_\mathrm{p}$ mutually orthogonal pilot sequences satisfying $\boldsymbol{\phi}_t^H \boldsymbol{\phi}_{t'} = \tau_\mathrm{p} \delta_{t,t'}$. In this paper, we consider a dense network scenario characterized by $K \ge \tau_\mathrm{p}$, which necessitates pilot reuse. Accordingly, let $t_k \in \{1,\ldots,\tau_\mathrm{p}\}$ denote the index of the pilot sequence assigned to UE~$k$.

To manage the MUI effectively, we adopt the RSMA scheme during the downlink data transmission phase. Specifically, the message of each UE is split into common and private parts. The former are aggregated into a shared common stream decoded by all UEs, while the latter are mapped to dedicated private streams for individual decoding.

\subsection{Pilot Assignment Strategy}
We establish the pilot indices $\{t_k\}_{k=1}^K$ using a low-complexity greedy assignment algorithm detailed in~\cite{2020TCOM_Bjorson, unsupDNN}. The procedure is executed as follows:

Initially, the first $\tau_\mathrm{p}$ UEs are randomly assigned distinct orthogonal pilot sequences such that $t_i \neq t_j$ for $i \neq j$, where $i,j \in \{1, \dots, \tau_\mathrm{p}\}$. Subsequently, for each remaining UE~$k$, where $k \in \{\tau_p+1, \dots, K\}$, the system first identifies the \emph{Master} AP that provides the strongest LSF gain to UE~$k$. The index of the \emph{Master} AP is given by
\begin{equation}
    l^* = \mathop{\arg\max}_{l \in \{1,\dots,L\}} \beta_{k,l}.
\end{equation}

Next, the \emph{Master} AP $l^*$ selects the pilot index $\tau^*$ for UE~$k$ that minimizes the potential pilot contamination from the set of previously assigned $(k-1)$ UEs. This selection criterion is formulated as
\begin{equation}
    \tau^* = \mathop{\arg\min}_{t \in \{1,\dots,\tau_\mathrm{p}\}} \sum_{\substack{i=1 \\ t_i = t}}^{k-1} \beta_{i,l^*}.
\end{equation}

We then set $t_k = \tau^*$. This procedure is repeated sequentially until all $K$ UEs are assigned pilot indices. This strategy effectively reduces the likelihood of assigning identical pilots to UEs with strong spatial correlation, thereby minimizing the dominant interference components.

\subsection{Uplink Signal Model and Channel Estimation}
During the uplink pilot training phase, all $K$ UEs simultaneously transmit their assigned pilot sequences to the APs. The received pilot signal matrix $\mathbf{Y}_{\mathrm{p},l} \in \mathbb{C}^{N \times \tau_\mathrm{p}}$ at AP~$l$ is modeled as
\begin{equation}
    \mathbf{Y}_{\mathrm{p},l} = \sum_{i=1}^{K} \sqrt{\eta_i} \mathbf{h}_{i,l} \boldsymbol{\phi}_{t_i}^T + \mathbf{N}_{\mathrm{p},l},
\end{equation}
where $\eta_i$ represents the uplink pilot power coefficient of UE~$i$, and $\mathbf{N}_{\mathrm{p},l} \in \mathbb{C}^{N \times \tau_\mathrm{p}}$ denotes the noise matrix with independent $\mathcal{CN}(0, \sigma^2)$ entries. 

To estimate the channel ${\mathbf{h}}_{k,l}$, AP~$l$ performs normalized despreading by projecting the received signal $\mathbf{Y}_{\mathrm{p},l}$ onto the selected pilot sequence $\boldsymbol{\phi}_{t_k}^*$. The despread pilot signal $\mathbf{y}_{t_k,l} \in \mathbb{C}^{N \times 1}$ is expressed as
\begin{equation}
    \mathbf{y}_{t_k,l} = \frac{1}{\sqrt{\tau_\mathrm{p}}} \mathbf{Y}_{\mathrm{p},l} \boldsymbol{\phi}_{t_k}^* = \sum_{\substack{i=1 \\ t_i = t_k}}^{K} \sqrt{\tau_\mathrm{p} \eta_{i}}\mathbf{h}_{i,l}+\mathbf{n}_{t_k,l},
\end{equation}
where $\mathbf{n}_{t_k,l} \sim \mathcal{CN}(\mathbf{0}, \sigma^2 \mathbf{I}_N)$ is the effective noise vector. As derived in~\cite{2024arXiv_Benjamin}, applying the standard MMSE estimator yields the channel estimate
\begin{equation}
    \hat{\mathbf{h}}_{k,l}
    = \sqrt{\tau_\mathrm{p} \eta_k}\,\mathbf{R}_{k,l}
    \left(
    \sum_{\substack{i=1 \\ t_i = t_k}}^{K}
    \tau_\mathrm{p} \eta_i \mathbf{R}_{i,l}
    + \sigma^2 \mathbf{I}_N
    \right)^{-1}
    \mathbf{y}_{t_k,l}.
\end{equation}
The estimated channel follows the distribution
\[
\hat{\mathbf{h}}_{k,l} \sim \mathcal{CN}\!\left(
\mathbf{0},\;
\tau_\mathrm{p} \eta_k\, \mathbf{R}_{k,l}
\mathbf{\Psi}_{k,l}^{-1}
\mathbf{R}_{k,l}
\right),
\]
where
\[
\mathbf{\Psi}_{k,l}
= \mathbb{E}\!\left\{\mathbf{y}_{t_k,l}\mathbf{y}_{t_k,l}^{H}\right\}
=
\sum_{\substack{i=1 \\ t_i=t_k}}^{K}
\tau_\mathrm{p} \eta_i \mathbf{R}_{i,l}
+ \sigma^2 \mathbf{I}_N
\]
denotes the covariance matrix of the despread pilot signal $\mathbf{y}_{t_k,l}$.

\subsection{Downlink Signal Model}
RSMA is employed for downlink data transmission. To balance the computational complexity and interference suppression capability, we employ maximum ratio precoding for the common stream and regularized zero-forcing for the private streams. Let $\bar{\mathbf{w}}_{\mathrm{c},l}$ and $\bar{\mathbf{w}}_{k,l}$ denote the unnormalized beamforming directions constructed based on local CSI at AP~$l$, which is given by
\begin{align}
    \bar{\mathbf{w}}_{\mathrm{c},l} &= \frac{1}{\sqrt{K}} \sum_{i=1}^{K} \hat{\mathbf{h}}_{i,l}, \\
    \bar{\mathbf{w}}_{k,l} &= \left( \sum_{i=1}^{K} \zeta_i \hat{\mathbf{h}}_{i,l} \hat{\mathbf{h}}_{i,l}^{{H}} + \sigma^2 \mathbf{I}_N \right)^{-1} \zeta_k \hat{\mathbf{h}}_{k,l},
\end{align}
where $\zeta_k$ represents the weight coefficient for UE~$k$. The final transmit beamforming vectors $\mathbf{w}_{\mathrm{c},l}$ and $\mathbf{w}_{k,l}$ are obtained by normalizing these directions to satisfy unit norm, i.e., $\|\mathbf{w}_{\mathrm{c},l}\|_{2} = \|\mathbf{w}_{k,l}\|_{2} = 1$.

Consequently, the transmitted signal vector $\mathbf{x}_l \in \mathbb{C}^{N \times 1}$ at AP~$l$ is expressed as
\begin{equation}
    \mathbf{x}_l = \sqrt{\rho_{\mathrm{c},l}}{\mathbf{w}}_{\mathrm{c},l} s_\mathrm{c} + \sum_{i = 1}^K \sqrt{\rho_{i,l}}{\mathbf{w}}_{il}s_i,
\end{equation}
where $s_\mathrm{c}, s_i \sim \mathcal{CN}(0,1)$ denote the normalized data symbols. The variables $\rho_{\mathrm{c},l}$ and $\rho_{i,l}$ represent the downlink transmit powers allocated by AP~$l$. For optimization convenience, we introduce the non-negative power coefficient vectors $\boldsymbol{\mu}_\mathrm{c} = [\mu_{\mathrm{c},1},\ldots,\mu_{\mathrm{c},L}]^{{T}}$ and $\boldsymbol{\mu}_{k} = [\mu_{k,1},\ldots,\mu_{k,L}]^{{T}}$, such that $\rho_{\mathrm{c},l} = \mu_{\mathrm{c},l}^{2}$ and $\rho_{k,l} = \mu_{k,l}^{2}$. The transmission at each AP satisfies the maximum power constraint $P_\mathrm{t}$, given by
\begin{equation}
    \rho_{\mathrm{c},l} + \sum_{i=1}^{K}\rho_{i,l} \leq P_\mathrm{t}, \quad \forall l \in \{1,\dots,L\}.
\end{equation}

By leveraging the TDD channel reciprocity, the received signal $y_{k} \in \mathbb{C}$ at UE~$k$ is given by
\begin{equation}
    y_{k} = \sum_{l=1}^{L} \mathbf{h}_{k,l}^{{H}} \mathbf{x}_l + n_k,
\end{equation}
where $n_k \sim \mathcal{CN}(0, \sigma^2)$ is the additive Gaussian noise at UE~$k$.

\subsection{Performance Metrics}
To facilitate a unified formulation of the SINR expressions, we introduce the stream type indicator $\alpha\in\{\mathrm{c},\mathrm{p}\}$ and let $\mathbf{w}_{k,l}^{\alpha}$ denote the beamforming vector at AP~$l$ for the stream of type $\alpha$ intended for UE~$k$. Specifically, for the common stream ($\alpha = \mathrm{c}$), the beamformer is shared by all UEs, i.e., $\mathbf{w}_{k,l}^\mathrm{c}=\mathbf{w}_{\mathrm{c},l},\forall i$. For the private stream ($\alpha= \mathrm{p}$), it corresponds to the user-specific beamformer, i.e., $\mathbf{w}_{k,l}^\mathrm{p}=\mathbf{w}_{k,l}$.

Based on these definitions, we utilize the use-and-then-forget (UatF) bound to derive the achievable rates. First, we define the effective channel gain provided by AP~$l$ to UE~$k$ for stream type $\alpha$ as
\begin{equation}
    a_{kl,\alpha} = \mathbb{E}\left\{\mathbf{h}_{k,l}^{{H}} \mathbf{w}_{k,l}^{\alpha} \right\}.
\end{equation}

By aggregating the gains from all $L$ APs, the effective channel gain vector $\mathbf{a}_{k,\alpha} \in \mathbb{R}^{L\times1}$ is formulated as
\begin{equation}
    \mathbf{a}_{k,\alpha} = \left[a_{k1,\alpha}, a_{k2,\alpha}, \dots, a_{kL,\alpha} \right]^{{T}}.
\end{equation}

Similarly, to characterize the interference power, we compute the interference correlation term between APs $l$ and $m$, arising from the stream of type $\alpha$ intended for UE~$i$, interfering with UE~$k$
\begin{equation}
    b_{ki,\alpha}^{lm} = \mathcal{R}\left(\mathbb{E}\left\{{\mathbf{h}}_{k,l}^H{\mathbf{w}}_{i,l}^{\alpha}{(\mathbf{w}}_{i,m}^{\alpha})^H{\mathbf{h}}_{k,m} \right\} \right),
\end{equation}
where $b_{ki,\alpha}^{lm}$ denotes the $(l,m)$-th entry of the interference correlation matrix ${\mathbf{B}}_{ki,\alpha} \in \mathbb{R}^{L \times L}$.

Consequently, the SINRs of the common and private streams at UE~$k$ are given by
\begin{align}
    \mathrm{SINR}_{k,\mathrm{c}}({\boldsymbol{\mu}}_\mathrm{c}) &= \frac{({\mathbf{a}}_{k,\mathrm{c}}^T{\boldsymbol{\mu}}_\mathrm{c})^2} {I_{k,\mathrm{c}}+\sum_{i=1}^K{\boldsymbol{\mu}}_{i}^T\mathbf{B}_{ki,\mathrm{p}}{\boldsymbol{\mu}}_{i}+\sigma^2}, \label{SINRcom}\\
    \mathrm{SINR}_{k,\mathrm{p}}({\boldsymbol{\mu}}_{k}) &= \frac{({\mathbf{a}}_{k,\mathrm{p}}^T{\boldsymbol{\mu}}_{k})^2}{\sum_{i=1}^K{\boldsymbol{\mu}}_{i}^T\mathbf{B}_{ki,\mathrm{p}}{\boldsymbol{\mu}}_{i}-({\mathbf{a}}_{k,\mathrm{p}}^T{\boldsymbol{\mu}}_{k})^2+\sigma^2}, \label{SINRpri}
\end{align}
where $I_{k,\mathrm{c}} = {\boldsymbol{\mu}}_\mathrm{c}^T\mathbf{B}_{kk,\mathrm{c}}{\boldsymbol{\mu}}_\mathrm{c}-({\mathbf{a}}_{k,\mathrm{c}}^T{\boldsymbol{\mu}}_\mathrm{c})^2$ represents the interference power caused by the channel estimation uncertainty of the common stream.

Finally, the achievable rate is defined as the sum of the minimum common stream rate and all private stream rates, which is given by
\begin{align}\label{SE}
\mathrm{SE}
&= \frac{\tau_{d}}{\tau_\mathrm{c}}
\bigg(
\min_{k}\,
\left\{\log_{2}\left(1+\mathrm{SINR}_{k,\mathrm{c}}(\boldsymbol{\mu}_\mathrm{c})\right) \right\} \notag \\ % 这里加上 \notag 
&\quad +\sum_{k=1}^{K}
\log_{2}\left(1+\mathrm{SINR}_{k,\mathrm{p}}(\boldsymbol{\mu}_{k})\right)
\bigg).
\end{align}

\section{Problem Formulation and Solution Methodology}
In this section, we formulate the downlink PA problem for the common and private streams. The objective is to maximize the achievable rate by optimizing the power coefficients \(\{ \rho_{k,l}\}\) and $\{\rho_{\mathrm{c},l}\}$, subject to per-AP power constraints. To tackle this problem, we derive a conventional iterative algorithm combining WMMSE and ADMM. This numerical approach serves as a performance upper bound for assessing the proposed GNN-based framework in the subsequent sections.

\subsection{Problem Formulation}
Mathematically, the sum SE maximization problem is formulated as
\begin{align}\label{OriOpt}
    \max_{\boldsymbol{\mu}_\mathrm{c}, \{\boldsymbol{\mu}_k\}} \quad & \frac{\tau_{d}}{\tau_\mathrm{c}} \left( \min_{k} R_{\mathrm{c},k} + \sum_{k=1}^{K} R_{\mathrm{p},k} \right) \\ 
    \textrm{s.t.} \quad & \mu_{\mathrm{c},l}^2 + \sum_{k=1}^{K}\mu_{k,l}^2 \leq P_\mathrm{t}, \quad \forall l \in \{1,\dots,L\},\tag{\ref{OriOpt}a}\label{const:power}
\end{align}
where the achievable rates $R_{\mathrm{c},k} = \log_{2}(1+\mathrm{SINR}_{k,\mathrm{c}})$ and $R_{\mathrm{p},k} = \log_{2}(1+\mathrm{SINR}_{k,\mathrm{p}})$ are determined by the SINRs as defined in \eqref{SINRcom} and \eqref{SINRpri}.

\subsection{WMMSE-ADMM Algorithm}\label{S_WMMSEADMM}
To solve the non-convex optimization problem formulated in \eqref{OriOpt}, we first exploit the WMMSE equivalence to obtain a tractable reformulation. Subsequently, we develop an ADMM-based procedure to solve the reformulated subproblems efficiently, which enables closed-form updates for all optimization variables and is theoretically guaranteed to converge to a stationary point.

For UE~$k$, let $v_{k,\mathrm{c}}$ and $v_{k,\mathrm{p}}$ denote the scalar equalizers for the common and private streams, respectively. The corresponding mean square errors (MSEs) are defined as
\begin{equation}\label{eck}
\begin{aligned}
e_{k,\mathrm{c}}
= {}& v_{k,\mathrm{c}}^{2}\!\left(
\boldsymbol{\mu}_\mathrm{c}^{{T}}\mathbf{B}_{kk,\mathrm{c}}\boldsymbol{\mu}_\mathrm{c}
+ \sum_{i=1}^{K}\boldsymbol{\mu}_{i}^{{T}}\mathbf{B}_{ki,\mathrm{p}}\boldsymbol{\mu}_{i}
+ \sigma^{2}
\right) \\
& {}- 2v_{k,\mathrm{c}}\,\mathbf{a}_{k,\mathrm{c}}^{{T}}\boldsymbol{\mu}_\mathrm{c}
+ 1,
\end{aligned}
\end{equation}
\begin{equation}\label{epk}
\begin{aligned}
e_{k,\mathrm{p}}
= {}& v_{k,\mathrm{p}}^{2}\!\left(
\sum_{i=1}^{K}
\boldsymbol{\mu}_{i}^{{T}}\mathbf{B}_{ki,\mathrm{p}}\boldsymbol{\mu}_{i}
+ \sigma^{2}
\right) \\
& {}- 2v_{k,\mathrm{p}}\,\mathbf{a}_{k,\mathrm{p}}^{{T}}\boldsymbol{\mu}_{k}
+ 1.
\end{aligned}
\end{equation}

By introducing auxiliary weights $w_{k,\alpha} > 0$, the equivalent WMMSE optimization problem can be formulated as
% 将 \small 放在环境外面
{\small
\begin{align}\label{Opt2}
    % 使用 \substack 让长下标自动换行，或者直接写
    \min_{\substack{\boldsymbol{\mu}_\mathrm{c},\, \{\boldsymbol{\mu}_{k}\}, \\ \{w_{k,\alpha},\, v_{k,\alpha}\}}} \quad &
    \sum\limits_{k = 1}^{K} \sum_{\alpha \in \{\mathrm{c},\mathrm{p}\}} \left(w_{k,\alpha}e_{k,\alpha} - \ln w_{k,\alpha}\right) \\
    \text{s.t.} \quad &  \mu_{\mathrm{c},l}^2 + \sum\limits_{k=1}^{K} \mu_{k,l}^2 \leq P_\mathrm{t}, \quad \forall l \in \{1, \dots, L\}.\tag{\ref{Opt2}a} \label{eq:const4}
\end{align}
}% 结束 small

Given fixed power coefficients $\boldsymbol{\mu}_{k}$, $\boldsymbol{\mu}_\mathrm{c}$ and auxiliary weights $w_{k,\alpha}$, the optimal equalizer $v_{k,\alpha}^{\mathrm{opt}}$ are obtained by setting the first-order derivative of \eqref{Opt2} to zero, yielding the closed-form updates as follows
\begin{equation}\label{vkcopt}
    v_{k,\mathrm{c}}^{\mathrm{opt}} = \frac{{\mathbf{a}_{k,\mathrm{c}}^T\boldsymbol{\mu}_\mathrm{c}}}{\tilde{v}_{k,\mathrm{c}}},
\end{equation}
\begin{equation}\label{vkpopt}
    v_{k,\mathrm{p}}^{\mathrm{opt}} = \frac{{\mathbf{a}_{k,\mathrm{p}}^T\boldsymbol{\mu}_{k}}}{\tilde{v}_{k,\mathrm{p}}},
\end{equation}
where
\begin{equation}
    \tilde{v}_{k,\mathrm{c}} = \boldsymbol{\mu}_\mathrm{c}^T \mathbf{B}_{kk,\mathrm{c}}\boldsymbol{\mu}_\mathrm{c}+\sum_{i= 1}^{K}{\boldsymbol{\mu}_{i}^T\mathbf{B}_{ki,\mathrm{p}}\boldsymbol{\mu}_{i}}+\sigma^2,
\end{equation}
\begin{equation}
    \tilde{v}_{k,\mathrm{p}} = \sum_{i= 1}^{K}{\boldsymbol{\mu}_{i}^T\mathbf{B}_{ki,\mathrm{p}}\boldsymbol{\mu}_{i}}+\sigma^2.
\end{equation}

Substituting \eqref{vkcopt} and \eqref{vkpopt} into \eqref{eck} and \eqref{epk}, we obtain the minimum MSE as
\begin{equation}
    e_{k,\alpha}^{\mathrm{MMSE}} = \frac{1}{1+\mathrm{SINR}_{k,\alpha}},
\end{equation}

Consequently, the optimal MSE weights $w_{k,\alpha}^{\mathrm{opt}}$ are updated as
\begin{equation}\label{wktopt}
    w_{k,\alpha}^{\mathrm{opt}} = e_{k,\alpha}^{-1}.
\end{equation}

The WMMSE procedure alternates among updating the equalizers ${v_{k,\alpha}}$, the weights ${w_{k,\alpha}}$, and the power coefficients ${\boldsymbol{\mu}_{k},\boldsymbol{\mu}_\mathrm{c}}$. Specifically, with fixed equalizers and weights, the power-update subproblem is equivalent to solving the following quadratically constrained quadratic program (QCQP)

\begin{align}\label{Opt3}
\operatorname*{min}_{\boldsymbol{\mu}} \quad & f({\boldsymbol{\mu}})\\
\text{s.t.} \quad & \mu_{\mathrm{c},l}^2 + \sum_{k=1}^{K} \mu_{k,l}^2 \leq P_\mathrm{t}, \quad \forall l, \tag{\ref{Opt3}a} \label{eq:const3}
\end{align}
where
\begin{equation}\label{fmu}
    f({\boldsymbol{\mu}}) = w_{k,\mathrm{c}}e_{k,\mathrm{c}} + \sum_{k = 1}^{K}w_{k,\mathrm{p}}e_{k,\mathrm{p}},
\end{equation}
and $\boldsymbol{\mu}^T = [\boldsymbol{\mu_{1}}^T,\boldsymbol{\mu_{2}}^T,\cdots,\boldsymbol{\mu_\mathrm{c}}^T]\in\mathbb{R}^{(K+1)L}$. The formulated subproblem \eqref{Opt3} is convex and can be efficiently solved by the ADMM-based procedure summarized in Algorithm~\ref{alg:wmmse-admm}.
\begin{algorithm}[t]
\caption{WMMSE--ADMM Algorithm for PA}
\label{alg:wmmse-admm}
\begin{algorithmic}[1]
\State \textbf{Initialization:} Set iteration index $r=0$. Initialize $v_{k,\alpha}^{(0)}$, $w_{k,\alpha}^{(0)}$, and $\boldsymbol{\mu}^{(0)}$ for all $k, \alpha$.
\Repeat
    \State \textbf{Step 1: Update Receive Filters}
    \State \quad Calculate $v_{k,\alpha}^{(r+1)}$ according to \eqref{vkcopt} and \eqref{vkpopt}.

    \State \textbf{Step 2: Update MSE Weights}
    \State \quad Calculate $w_{k,\alpha}^{(r+1)}$ according to \eqref{wktopt}.

    \State \textbf{Step 3: Update Power Allocation (ADMM)}
    \State \quad Solve the QCQP problem \eqref{Opt3} to obtain $\boldsymbol{\mu}^{(r+1)}$.

    \State $r \leftarrow r+1$.
\Until{convergence of $\boldsymbol{\mu}^{(r)}$ or the objective value $f(\boldsymbol{\mu}^{(r)})$}
\end{algorithmic}
\end{algorithm}

The computational complexity of the proposed WMMSE-ADMM algorithm is dominated by the iterative updates of the optimization variables.
In each outer iteration, updating the scalar equalizers $\{v_{k,\alpha}\}$ and weights $\{w_{k,\alpha}\}$ entails calculating the SINR values, which involves matrix-vector multiplications with a complexity of $\mathcal{O}(K^2 L^2)$.
For the PA update via ADMM, the computational bottleneck lies in the primal $\boldsymbol{\mu}$-update step.
Specifically, solving the unconstrained quadratic subproblem requires determining the Cholesky factorization of a large-scale matrix with dimension $(K+1)L \times (K+1)L$, resulting in a complexity of $\mathcal{O}(K^3 L^3)$.
Although the subsequent projection step onto the per-AP power constraints can be decomposed across APs with a lower linear complexity of $\mathcal{O}(LK)$, the overall computational burden remains dictated by the matrix operations.
Consequently, the total computational complexity of the WMMSE-ADMM algorithm is approximately $\mathcal{O}\big(I_{\mathrm{out}} (K^3 L^3 + I_{\mathrm{in}} K^2 L^2)\big)$, where $I_{\mathrm{out}}$ and $I_{\mathrm{in}}$ denote the number of outer WMMSE iterations and inner ADMM iterations, respectively.
This high polynomial complexity with respect to the system size $L$ and $K$ renders the WMMSE-ADMM approach computationally prohibitive for real-time applications in large-scale CF-mMIMO systems.

\section{Proposed Scalable GNN Framework}
\begin{figure*}[!t]
\centering
\includegraphics[width=\textwidth]{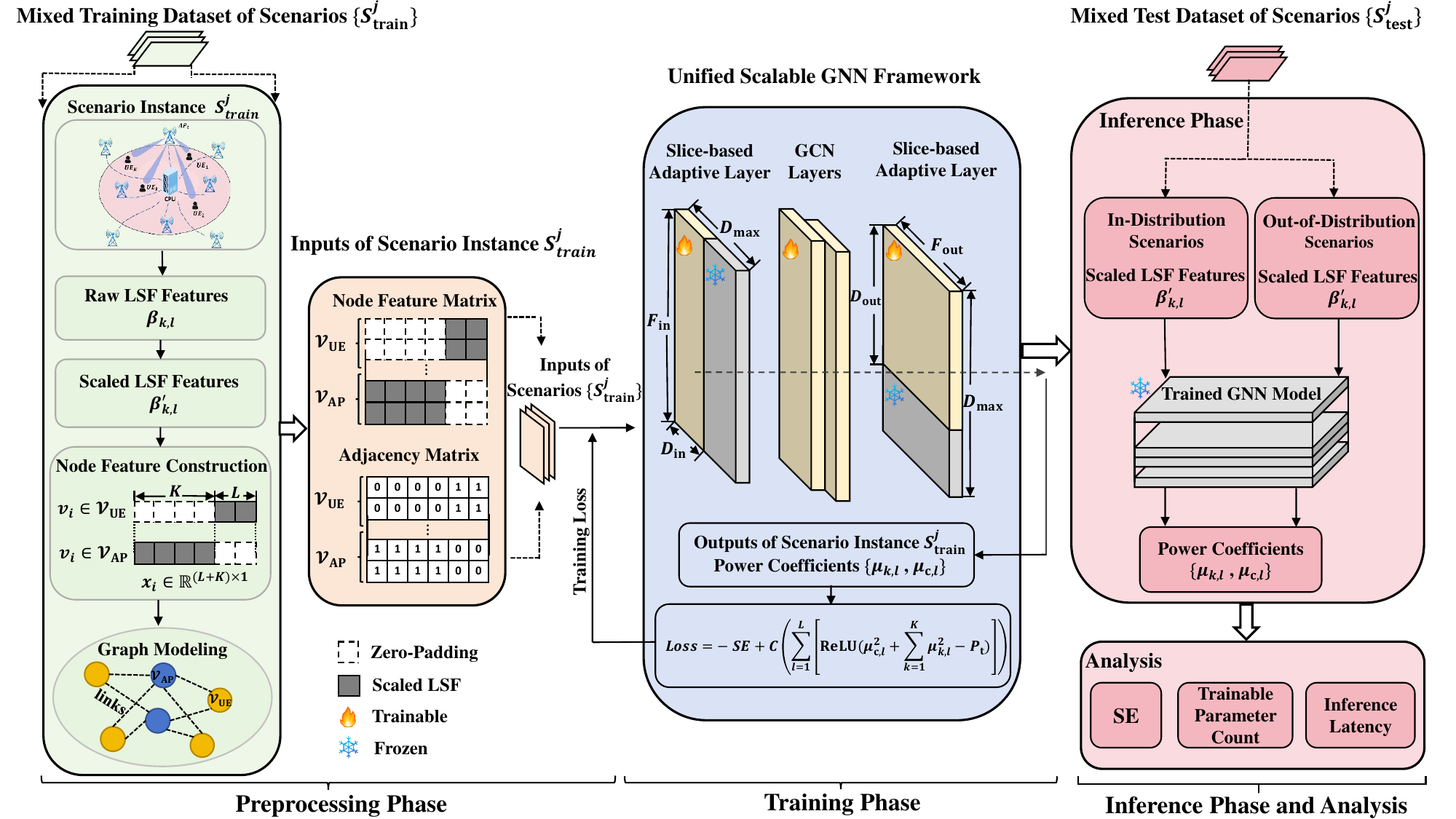}
\caption{A schematic overview of the proposed GNN framework}
\label{fig_2}
\end{figure*}
Although the WMMSE-ADMM algorithm derived in Section~\ref{S_WMMSEADMM} provides a near-optimal solution for the problem in \eqref{OriOpt}, its practical deployment is hindered by the prohibitive overhead arising from iterative matrix operations. To overcome these latency and scalability bottlenecks, we propose a scalable unsupervised GNN-based PA framework deployed on the CPU of an edge server. This centralized deployment enables real-time inference by shifting the heavy optimization burden to offline training, allowing the CPU to efficiently compute and distribute PA coefficients to individual APs using aggregated LSF information. Crucially, unlike fixed-size neural networks, our framework features a topology-adaptive architecture that decouples model complexity from the system size.

\subsection{Graph Modeling and Feature Construction}\label{S_graphmodel}
The RS-CF-mMIMO system is naturally modeled as a bipartite graph $\mathcal{G}=(\mathcal{V}, \mathcal{E})$. The node set $\mathcal{V}=\mathcal{V}_{\text{UE}} \cup \mathcal{V}_{\text{AP}}$ comprises $K$ UE nodes and $L$ AP nodes. Specifically, we assign indices $\{1, \dots, K\}$ to $\mathcal{V}_{\text{UE}}$ and $\{K+1, \dots, K+L\}$ to $\mathcal{V}_{\text{AP}}$. The edge set $\mathcal{E}$ represents the active communication links between APs and UEs. We utilize the LSF coefficients $\{\beta_{k,l}\}$ as raw node features due to their stability and dominance in PA decisions. To mitigate the large dynamic range of channel gains, the LSF values are normalized as
\begin{equation}\label{LSF}
    \beta^\prime_{k,l} = \sqrt{P_t}\frac{(\beta_{k,l})^\xi}{\sum_{m = 1}^K(\beta_{m,l})^\xi},
\end{equation}
where $\xi$ is a tunable exponent. Leveraging this topology, we construct a raw feature vector $\mathbf{x}_i \in \mathbb{R}^{(L+K) \times 1}$ for each node $v_i \in \mathcal{V}$. To preserve the distinct roles and connectivity patterns of APs and UEs, the feature vectors are structured according to the node type
\begin{equation}
\mathbf{x}_i = 
\begin{cases} 
[\beta_{i,1}^{\prime}, \dots, \beta_{i,L}^{\prime}, \mathbf{0}_{1 \times K}]^T, & \text{if } v_i \in \mathcal{V}_{\text{UE}}, \\
[\mathbf{0}_{1 \times L}, \beta_{1,l}^{\prime}, \dots, \beta_{K,l}^{\prime}]^T, & \text{if } v_i \in \mathcal{V}_{\text{AP}}, 
\end{cases} 
\label{eq:feature_vector}
\end{equation}
where for any AP node, the corresponding AP index is mapped by $l = i - K$. Crucially, the zero blocks $\mathbf{0}_{1\times K}$ and $\mathbf{0}_{1\times L}$ serve as structural indicators that explicitly encode the bipartite constraints, i.e., the absence of direct UE-UE or AP-AP links. 

\subsection{Scalable Architecture Design}
The schematic overview of the proposed GNN framework is illustrated in Fig.~\ref{fig_2} with specific layer configurations detailed in Table~\ref{table_gnn}. The architecture primarily comprises three functional modules: a slice-based adaptive embedding layer serving as the scalable interface, a multi-layer graph convolutional network (GCN) backbone equipped with residual connections for deep feature extraction, and an adaptive projection layer that maps the refined node embeddings onto the topology-dependent PA coefficients.

\begin{table}[!t]
\caption{Layer Configurations of the Proposed Scalable GNN Framework}
\label{table_gnn}
\centering
\begin{tabular}{cccc}
\toprule
\textbf{Layer} & \textbf{Type} & \textbf{Output Dimension} & \textbf{Activation} \\
\midrule
Input   & Raw Feature                 & $K+L$        & -- \\
Layer 1 & Adaptive Embedding          & $48$         & Linear \\
Layer 2 & GCN Backbone ($t=1$)        & $64$         & ReLU \\
Layer 3 & GCN Backbone ($t=2$)        & $128$        & ReLU \\
Layer 4 & MLP                         & $64$         & ReLU \\
Output  & Adaptive Projection         & $(K+1)L$     & Sigmoid \\
\bottomrule
\end{tabular}
\end{table}

\subsubsection{Slice-Based Adaptive Embedding}
\begin{figure}[!t]
\centering
\includegraphics[width=3 in]{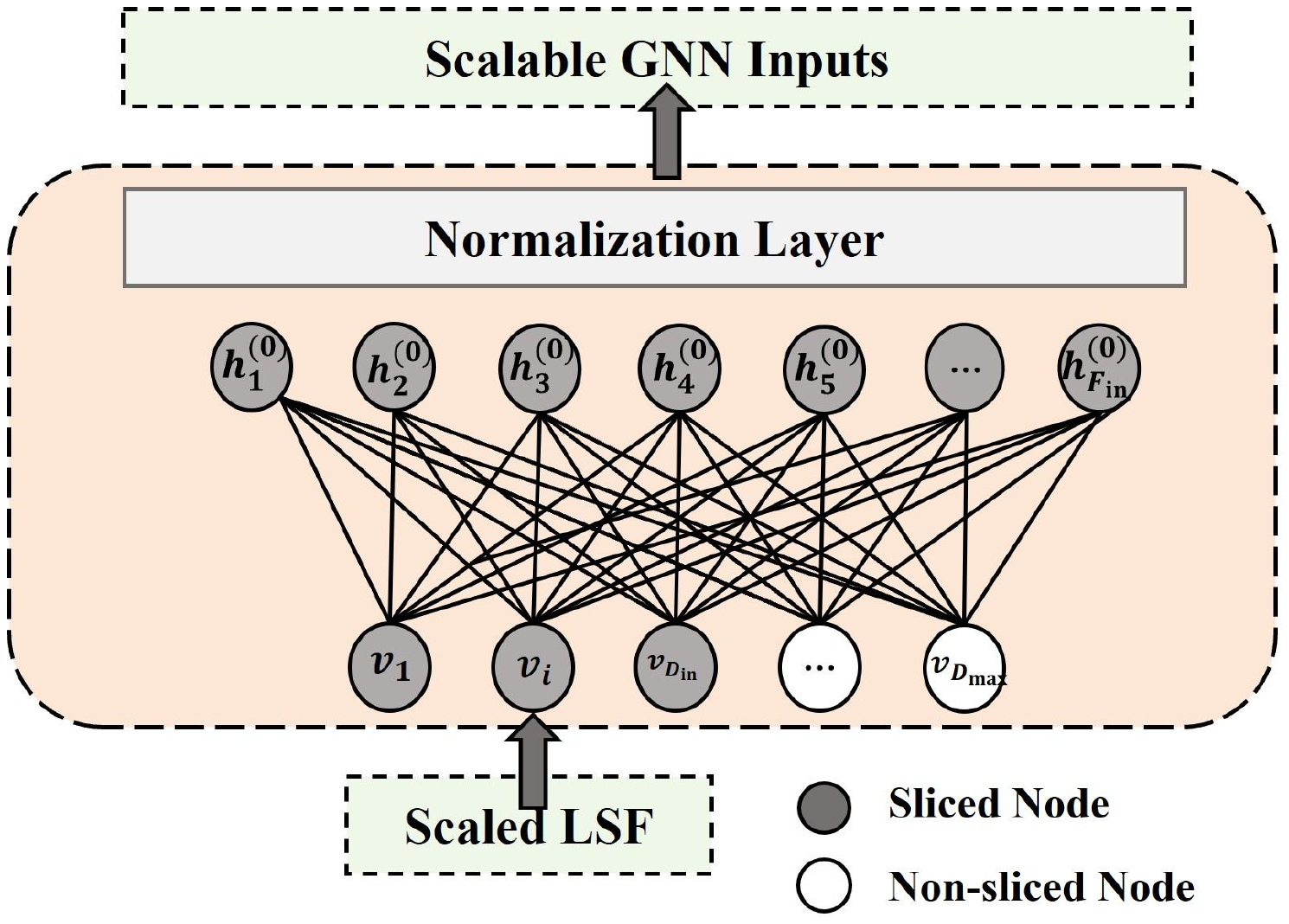}
\caption{Illustration of the adaptive node embedding layer mechanism.}
\label{setting}
\end{figure}
As the first layer of the proposed scalable architecture, the slice-based adaptive embedding layer in Fig.~\ref{setting} provides a topology-adaptive interface that allows a single unified GNN to operate across CF deployments with varying $(L,K)$. Specifically, we introduce a global trainable weight pool $\mathbf{W}_{\mathrm{in}}\in\mathbb{R}^{F_\mathrm{in}\times D_\mathrm{max}}$, where $F_\mathrm{in}$ denotes the target latent dimension and $D_\mathrm{max}=L_\mathrm{max}+K_\mathrm{max}$ represents the maximum system capacity, which is set to 64 in our simulations. For a deployment with input dimension $D_{\mathrm{in}}=L+K$, the layer activates the corresponding prefix sub-matrix $\mathbf{W}_{\mathrm{in}}[:,1\!:\!D_{\mathrm{in}}]$ with dimension $F_\mathrm{in}\times D_{\mathrm{in}}$ and computes the initial node embedding as
\begin{equation}
\mathbf{h}_i^{(0)}=\mathbf{W}_{\mathrm{in}}[:,1\!:\!D_{\mathrm{in}}]\mathbf{x}_i,
\end{equation}
thereby mapping variable-dimensional topology features into a fixed-dimensional latent space of size $F_\mathrm{in}$ prior to message passing.

In conventional padding-based designs, each $\mathbf{x}_i$ is first expanded to the maximum dimension $D_\mathrm{max}$, and the embedding is computed using a fixed $F_\mathrm{in}\times F_\mathrm{max}$ transformation. In contrast, our layer performs the multiplication with an effective matrix of size $F_\mathrm{in}\times D_{\mathrm{in}}$, so that the embedding cost scales with the actual deployment size as $\mathcal{O}(F_\mathrm{in} D_{\mathrm{in}})=\mathcal{O}(F_\mathrm{in}(L+K))$. Consequently, the proposed embedding avoids redundant computations and enables efficient real-time inference in dynamic CF-mMIMO deployments.

\subsubsection{Graph Convolutional Processing}
To capture high-order interference dependencies embedded in the network topology, the initial node embeddings are processed by a stack of $T$ GCN layers. We employ a spectral graph convolution operator with symmetric normalization to ensure stable and degree-unbiased feature aggregation. Specifically, the feature update for node $v_i$ at layer $t\in\{1,\dots,T\}$ is given by
\begin{equation}
    \mathbf{h}_i^{(t)} = \sigma \left(
    \mathbf{W}^{(t)} \sum_{v_j \in \mathcal{N}(v_i)} c_{i,j} \mathbf{h}_j^{(t-1)}
    + \mathbf{h}_i^{(t-1)}
    \right),
    \label{eq:gcn_update}
\end{equation}
where $\mathcal{N}(v_i)$ denotes the set of first-order neighbors of node $v_i$ and $\mathbf{W}^{(t)}$ is the trainable weight matrix at layer $t$. $c_{i,j}=1/\sqrt{|\mathcal{N}(v_i)|\,|\mathcal{N}(v_j)}|$ and $|\mathcal{N}(\cdot)|$ denotes the cardinality of the corresponding neighbor set. The residual connection $\mathbf{h}_i^{(t-1)}$ is introduced to mitigate over-smoothing and facilitate gradient propagation during training, while $\sigma(\cdot)$ denotes the ReLU activation function.

By iteratively applying \eqref{eq:gcn_update}, the node representations progressively integrate local channel characteristics with broader topological interference information. Specifically, in a two-layer architecture ($T=2$), the initial message-passing iteration ($t=1$) aggregates immediate neighborhood features, directly encoding the spatial coupling between UEs and APs. The subsequent iteration ($t=2$) expands the receptive field to two-hop neighbors, systematically resolving the overlapping interference clusters among UEs sharing the same or adjacent APs. By projecting these topological dependencies into high-dimensional embeddings, the GCN implicitly characterizes complex MUI boundaries without requiring instantaneous small-scale fading, thereby establishing the essential interference awareness requisite for subsequent RSMA power allocation.

\subsubsection{Slice-based Adaptive Projection}
Following the GCN backbone, an adaptive projection layer maps the refined node embeddings of dimension $F_\mathrm{out}$ to the topology-dependent power coefficient vector with dimension $D_{\mathrm{out}}=(K+1)L$. The resulting projection output is passed through a sigmoid activation and scaled by $P_{\mathrm{t}}$, producing non-negative coefficients that satisfy the per-AP transmit power constraints.

\subsection{Scalability Analysis and Boundary Conditions}\label{sec:boundary}
For ultra-large-scale deployments in which the required dimension $D_{\mathrm{in}}$ exceeds $D_{\mathrm{max}}$, processing the entire monolithic topology with a single direct slice is structurally infeasible. To extend the framework’s capability to arbitrarily large deployments, we integrate a user-centric overlapping clustering strategy that decomposes the global topology into size-bounded sub-graphs. To execute this decomposition, we first construct a sparse UE-AP topology based on the dominant large-scale channel conditions, following the procedures as follows:

Each UE firstly associates with its $Q_\mathrm{AP}$ strongest candidate APs, defined by $\mathcal{A}_k = \mathrm{TopQ}_{l}(\beta_{k,l})$ with $|\mathcal{A}_k| = Q_\mathrm{AP}$. Furthermore, to regulate localized graph density, we define $\mathcal{U}_l$ as the set of UEs associated with AP $l$, where each AP retains at most $Q_\mathrm{UE}$ UEs with the highest channel gains, i.e., $|\mathcal{U}_l| \le Q_\mathrm{UE}$. Here, $Q_\mathrm{AP}$ and $Q_\mathrm{UE}$ serve as structural constraints to govern local graph density rather than training hyperparameters.

The global network is subsequently partitioned into overlapping sub-graphs. Let $\mathcal{C}_i \subseteq \mathcal{V}_{\mathrm{UE}}$ denote the set of core UEs assigned to the $i$-th sub-graph. The corresponding sub-graph $\mathcal{S}_i$ is constructed as the $T$-hop neighborhood of these core UEs, i.e., $\mathcal{S}_i = \mathcal{N}_T(\mathcal{C}_i)$, where $\mathcal{N}_T(\cdot)$ denotes the set of all AP and UE nodes within a $T$-hop distance of the core UEs in the global graph, thereby encompassing all nodes participating in the $T$-layer GNN message-passing operations. The total dimension of each sub-graph is constrained by 
\begin{equation}
    D_i = |\mathcal{S}_i \cap \mathcal{V}_{\mathrm{AP}}| + |\mathcal{S}_i \cap \mathcal{V}_{\mathrm{UE}}| \le D_{\mathrm{max}}. 
\end{equation}

To verify the structural integrity of this decomposition, we define the global $T$-hop receptive field of UE $k$ as $\mathcal{R}_T(k) = \mathcal{N}_T(\{k\}; \mathcal{G})$. If UE $k$ is designated as a core user for sub-graph $\mathcal{S}_{c(k)}$, its structural context is perfectly preserved if $\mathcal{R}_T(k) \subseteq \mathcal{S}_{c(k)}$. We quantify the context preservation efficacy via:
\begin{equation}
    \rho_T = \frac{1}{|\mathcal{V}_{\mathrm{UE}}|} \sum_{k \in \mathcal{V}_{\mathrm{UE}}} \mathbb{I} \left\{ \mathcal{R}_T(k) \subseteq \mathcal{S}_{c(k)} \right\},
\end{equation}
where $\mathbb{I}\{\cdot\}$ is the indicator function. A preservation ratio of $\rho_T = 100\%$ confirms that the essential interference context for each core UE is rigorously maintained. This mechanism guarantees that the pre-trained neural architecture generalizes seamlessly to arbitrary network scales without retraining, effectively circumventing the global dimensionality curse by shifting the scalability bottleneck from global counts to localized neighborhood density.

\subsection{Unsupervised Training Strategy}
The proposed framework is trained in an unsupervised manner. Based on the achievable-rate expressions derived in Section~\ref{S_system}, the SINRs of the common and private streams can be evaluated as explicit functions of the PA coefficients produced by the GNN. Since the common stream must be decoded by all UEs, its achievable rate is limited by the minimum user rate, whereas the private streams contribute additively to the overall throughput. Accordingly, the SE-based loss is defined as
\begin{align}\label{loss}
\mathcal{L}_\mathrm{SE}
=& - \min_{k} \, \log_2\!\left(1 + \mathrm{SINR}_{k,\mathrm{c}}\right) \notag \\
& - \frac{1}{K} \sum_{k=1}^{K} \log_2\!\left(1 + \mathrm{SINR}_{k,\mathrm{p}}\right) \notag \\
& + C \sum_{l=1}^{L} 
    \mathrm{ReLU}\!\left(
        \mu_{\mathrm{c},l}^2 +\sum_{k=1}^{K} \mu_{k,l}^2 - P_\mathrm{t}
    \right),
\end{align}

As the presence of the minimum operator in \eqref{loss} renders the objective non-differentiable, we approximate the minimum function using the LogSumExp (LSE) formulation
\begin{equation}
\min_k x_k \approx -\tau \ln\left( \sum_{k=1}^{K} \exp\left( -\frac{x_k}{\tau} \right) \right),
\end{equation}

The differentiable loss used for training is
\begin{align}\label{loss_approx}
\tilde{\mathcal{L}}_\mathrm{SE}
=&\;
-\tau \ln\!\left(
    \sum_{k=1}^{K}
    \exp\!\left(
        -\frac{\log_2\!\left(1+\mathrm{SINR}_{k,\mathrm{c}}\right)}{\tau}
    \right)
\right) \notag \\
& - \frac{1}{K} \sum_{k=1}^{K} \log_2\!\left(1+\mathrm{SINR}_{k,\mathrm{p}}\right) \notag \\
& + C \sum_{l=1}^{L}
    \mathrm{ReLU}\!\left(
        \mu_{\mathrm{c},l}^2 + \sum_{k=1}^{K} \mu_{k,l}^2 - P_\mathrm{t}
    \right),
\end{align}
where $\tau>0$ governs the trade-off between LogSumExp approximation accuracy and gradient smoothness, whereas $C$ dictates the enforcement of the power constraint. To establish a robust training foundation, we perform sensitivity analyses to identify stable operational regimes for these hyperparameters. 

Regarding the temperature $\tau$, empirical results show that while a lower value (e.g., $\tau=0.05$) marginally improves rate-approximation accuracy (by less than 0.3\%), it risks decelerating convergence due to gradient concentration. Conversely, $\tau=0.1$ maintains a superior balance, facilitating rapid convergence without compromising the SE. Regarding the penalty coefficient $C$, we observe that weak regularization ($C \le 0.01$) fails to effectively suppress power violations, whereas excessive penalization ($C \ge 0.2$) distorts the gradient descent trajectory, prematurely forcing the model into a feasibility-dominated regime. Consequently, $C=0.1$ is identified as the optimal configuration, effectively suppressing power violation rates to a negligible level ($<0.03\%$) while preserving the global optimization trajectory. Based on these validations, we adopt $\tau=0.1$ and $C=0.1$ across all deployments to ensure strict compliance with constraints and robust convergence.

\section{Simulation Results and Analysis}
\subsection{Simulation Setup}
% \begin{figure}[!t]
% \centering
% \includegraphics[width=2.5in]{Fig/SystemModel/Graph.pdf}
% \caption{A representative topology for one realization of the considered CF-mMIMO system.}
% \label{simset}
% \end{figure}

In this section, we present numerical results to evaluate the performance of the proposed scalable GNN PA framework. The simulations are conducted in an urban scenario within a square area of $1000\,\mathrm{m} \times 1000\,\mathrm{m}$. We consider networks comprising $L\in\{9,16,25,36\}$ distributed APs, each equipped with $N = 4$ antennas. The AP positions are fixed across channel realizations, assuming a wrapped-around topology to mitigate boundary effects. The APs jointly serve $K\in\{6,10,16,20\}$ single-antenna UEs that are randomly distributed within the coverage area. We simulate both orthogonal and non-orthogonal pilot assignment schemes over a bandwidth of $20\,\mathrm{MHz}$ with a receiver noise power of $-94$ dBm. The per-AP maximum downlink transmit power is $P_\mathrm{t} = 1\,\mathrm{W}$ and the UL pilot power is $\eta_i = 100$ mW. The LSF coefficients are generated based on the model in \cite{unsupDNN2025}
\begin{equation}
    \beta_{k,l} = -30.5-36.7\,log_{10}(\frac{d_{k,l}}{1\mathrm{m}}),
\end{equation}
where $d_{k,l}$ is the distance between UE~$k$ and AP~$l$. Key simulation parameters are summarized in Table~\ref{table2}.

\begin{table}[!t]
\caption{Simulation Parameters}
\label{table2}
\centering
% 建议去掉竖线，符合 IEEE Trans 风格
\begin{tabular}{cc}
\hline
\textbf{Parameter} & \textbf{Value} \\
\hline
Coverage Area  & $1000 \times 1000\,\mathrm{m}^2$   \\
System Bandwidth  & $20\,\mathrm{MHz}$   \\
Number of APs ($L$)   & $ \{9,16,25,36\}  $   \\
Number of UEs ($K$)   & $ \{6,10,16,20\}  $   \\
Antennas per AP ($N$)   & $4$  \\
Per-AP max downlink power ($P_\mathrm{t}$)  & $1\,\mathrm{W}$  \\
UL pilot power ($\eta_i$) & $100\,\mathrm{mW}$ \\
Coherence block length ($\tau_\mathrm{c}$) & 200\\
Receiver noise power  & $-94\,\mathrm{dBm}$ \\
Path-loss exponent ($\alpha$)  & $3.67$\\
LSF scaling exponent ($\xi$)  & $0.4$  \\
\hline
\end{tabular}
\end{table}

The proposed framework is implemented using PyTorch on an NVIDIA RTX 4090 GPU (24 GB). The models are trained using the Adam optimizer with a mini-batch size of $32$. The initial learning rate is set to $0.01$ and is dynamically decayed via a cosine-annealing scheduler. To maximize the system SE, the loss function defined in \eqref{loss_approx} is adopted. To comprehensively assess the proposed architecture, we conduct evaluations under two distinct training regimes: \textit{Scenario-Specific Training} and \textit{Mixed-Scenario Training}. We denote the datasets of diverse system configurations for training and testing as $\{S_\mathrm{train}^j\}$ and $\{S_\mathrm{test}^j\}$, respectively. Specific dataset configurations are detailed in the respective subsections.

% 1) \textbf{Scenario-Specific Training (Unscalable):} In this configuration, we evaluate the model's specialized performance for specific network sizes. A dedicated model is trained and tested solely on a single-scenario dataset comprising $20,000$ independent channel realizations generated for a fixed target $(L, K)$ pair. This approach assesses the architecture's capability when fully optimized for a known, static environment.

% 2) \textbf{Mixed-Scenario Training (Scalable):} In this configuration, we evaluate the model's generalization capability across varying network sizes. The scalable model is trained on a heterogeneous dataset aggregated from multiple distinct network configurations. Specifically, each configuration contributes $2,000$ samples. Consequently, the total dataset size $N_s$ ranges from $6,000$ to $16,000$, contingent upon the number of scenarios included in the specific experiment.

The performance of the proposed scalable GNN framework, denoted as \textbf{\emph{RSMA-GNN-U}}, is evaluated by comparing with a comprehensive set of seven distinct benchmark schemes, categorized as follows:
\begin{enumerate}[label=\textbf{(\roman*)}, leftmargin=*]
    \item \textbf{\emph{BC}:} A broadcasting strategy that serves as a performance floor.
    \item \textbf{\emph{RSMA-EP}}: An RSMA scheme employing uniform PA across all streams, serving as a specific lower bound for rate-splitting transmission.
    \item \textbf{\emph{RSMA-WMMSE}}: The WMMSE-ADMM algorithm formulated in Algorithm~\ref{alg:wmmse-admm}, which provides a performance upper bound.
    \item \textbf{\emph{RSMA-DNN-U}:} An adaptation of the DNN-based scheme~\cite{unsupDNN} for RSMA, serving as a benchmark to verify the architectural superiority of the proposed GNN.
    \item \textbf{\emph{SDMA-GNN-U}:} The proposed GNN framework adapted for SDMA, quantifying the explicit gain of RSMA.
    \item \textbf{\emph{SDMA-DNN-U}:} The conventional unsupervised DNN-based SDMA framework established in~\cite{unsupDNN}.
    \item \textbf{\emph{SDMA-DNN-S}:} A supervised DNN-based SDMA model~\cite{2023TWC_Bjornson} trained using labels generated by the WMMSE algorithm.
\end{enumerate}

This diverse selection of benchmarks enables a rigorous assessment of the proposed \emph{RSMA-GNN-U} framework, validating its advantages in terms of \textbf{SE}, \textbf{inference latency}, and \textbf{parameter count}.

\subsection{Offline Training Cost Analysis}
We first evaluate the dataset generation cost of the proposed framework. Unlike supervised methodologies that demand computationally prohibitive iterative solvers to yield ground-truth labels, the proposed unsupervised training set is constructed solely by sampling UE locations and computing the corresponding LSF coefficients and spatial correlation matrices. Empirically, generating a comprehensive dataset of $20,000$ samples on an NVIDIA RTX 4090 GPU takes only $114.4$ seconds (averaging $5.72$ ms per sample). This data acquisition cost is virtually negligible, as it circumvents the most time-consuming phase of conventional deep learning pipelines.
    
Furthermore, we quantify the offline training overhead of the unified mixed-scenario GNN. The model is trained across eight heterogeneous scenarios with $N=2{,}000$ samples per setup and a $0.9$ training split, yielding $14{,}400$ training samples per epoch. On the same GPU platform, the entire $20$-epoch training procedure, encompassing data loading, gradient updates, and validation, takes only $57.4$ minutes. It is imperative to note that this constitutes a strictly one-time offline cost.

For comparison, while the conventional WMMSE-ADMM algorithm avoids offline training, it still requires solving a complex iterative optimization problem for each new channel realization. Our measurements indicate that its average runtime is highly topology-dependent, escalating from $24.7$ seconds per sample in sparse networks (e.g., $L=16, K=6$) to $260.6$ seconds per sample in denser deployments (e.g., $L=16, K=16$). Consequently, the entire one-time offline training cost of the GNN is equivalent to merely $13$--$140$ WMMSE-ADMM executions. Moreover, evaluating the WMMSE-ADMM solver exclusively on the test split of the eight considered scenarios ($800$ samples) would demand approximately $31.4$ hours. In contrast, the GNN training concludes in under one hour. Once deployed, the GNN circumvents per-sample iterative optimization via an efficient direct forward pass. Therefore, the offline training overhead is highly affordable and is rapidly amortized by the substantial online computational savings.

\subsection{Performance Evaluation with Scenario-Specific Training}
We then evaluate the performance of the proposed framework in a scenario-specific setting. In this regime, a dedicated model is trained specifically for each fixed target $(L, K)$ pair utilizing orthogonal pilot assignment. The corresponding dataset comprises $20,000$ independent channel realizations, which are randomly split into training, validation, and test sets with a ratio of $8:1:1$. This configuration allows us to assess the architecture's peak capability when fully optimized for a known, static environment.

To rigorously validate the robustness of the proposed scheme under varying network loads, we define three representative scenarios with a fixed number of APs $L=16$: an underloaded scenario with $K=10$ ($K<L$); a fully loaded scenario with $K=16$ ($K=L$); and an overloaded scenario with $K=20$ ($K>L$). This setup aligns with the experimental conditions in \cite{2023TWC_Bjornson}. In the following analysis, we present the cumulative distribution function (CDF) of the downlink SE per UE. Additionally, the training and validation loss curves are provided to illustrate the convergence behavior.

\subsubsection{SE performance}
\begin{figure*}[!t]
\centering

\begin{minipage}[b]{0.32\textwidth} 
    \centering
    \includegraphics[width=\textwidth]{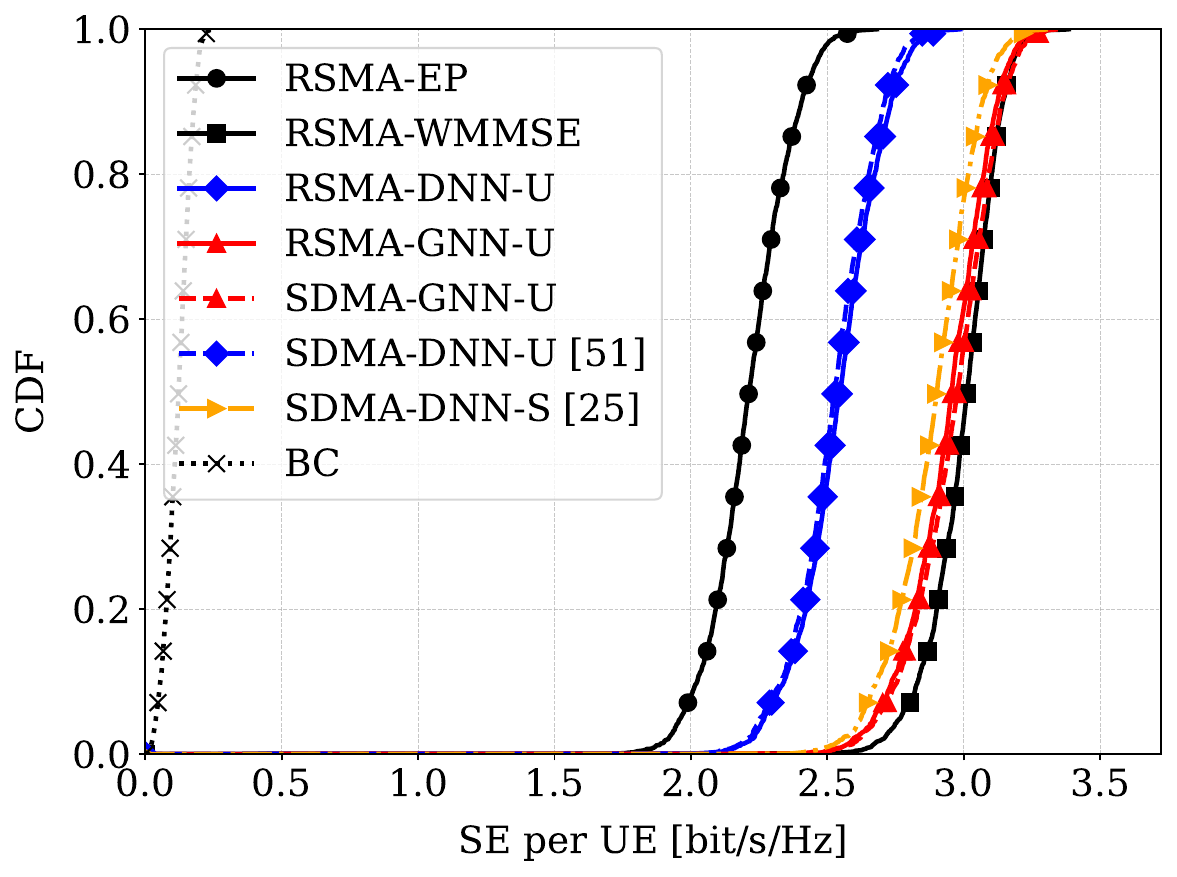} 
    \par\vspace{0.5em}
    \footnotesize (a) Underloaded: $K=10$ UEs
   \label{fig:se_cdf_a}
\end{minipage}
\hspace{-3mm} 
\begin{minipage}[b]{0.32\textwidth}
    \centering
    \includegraphics[width=\textwidth]{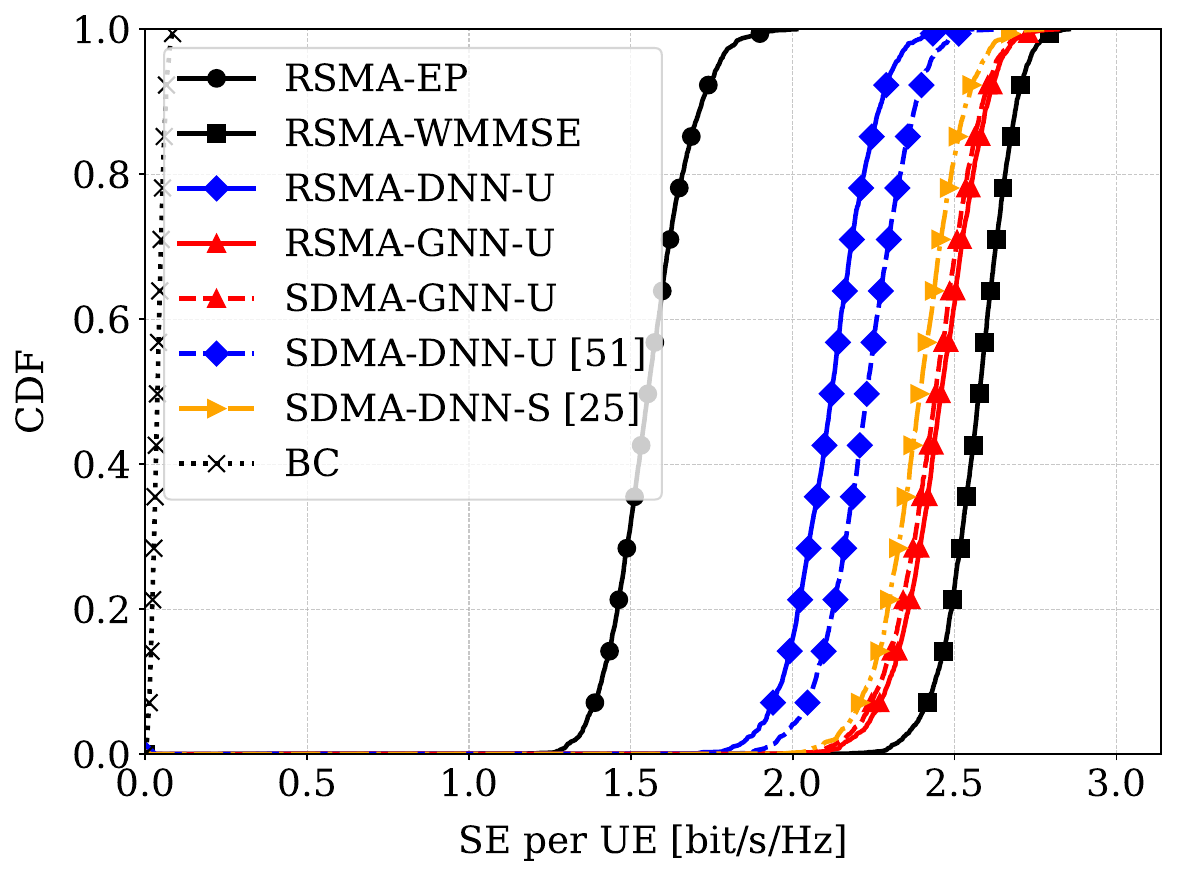}
    \par\vspace{0.5em}
    \footnotesize (b) Fully Loaded: $K=16$ UEs
    \label{fig:se_cdf_b}
\end{minipage}
\hspace{-3mm} 
\begin{minipage}[b]{0.32\textwidth}
    \centering
    \includegraphics[width=\textwidth]{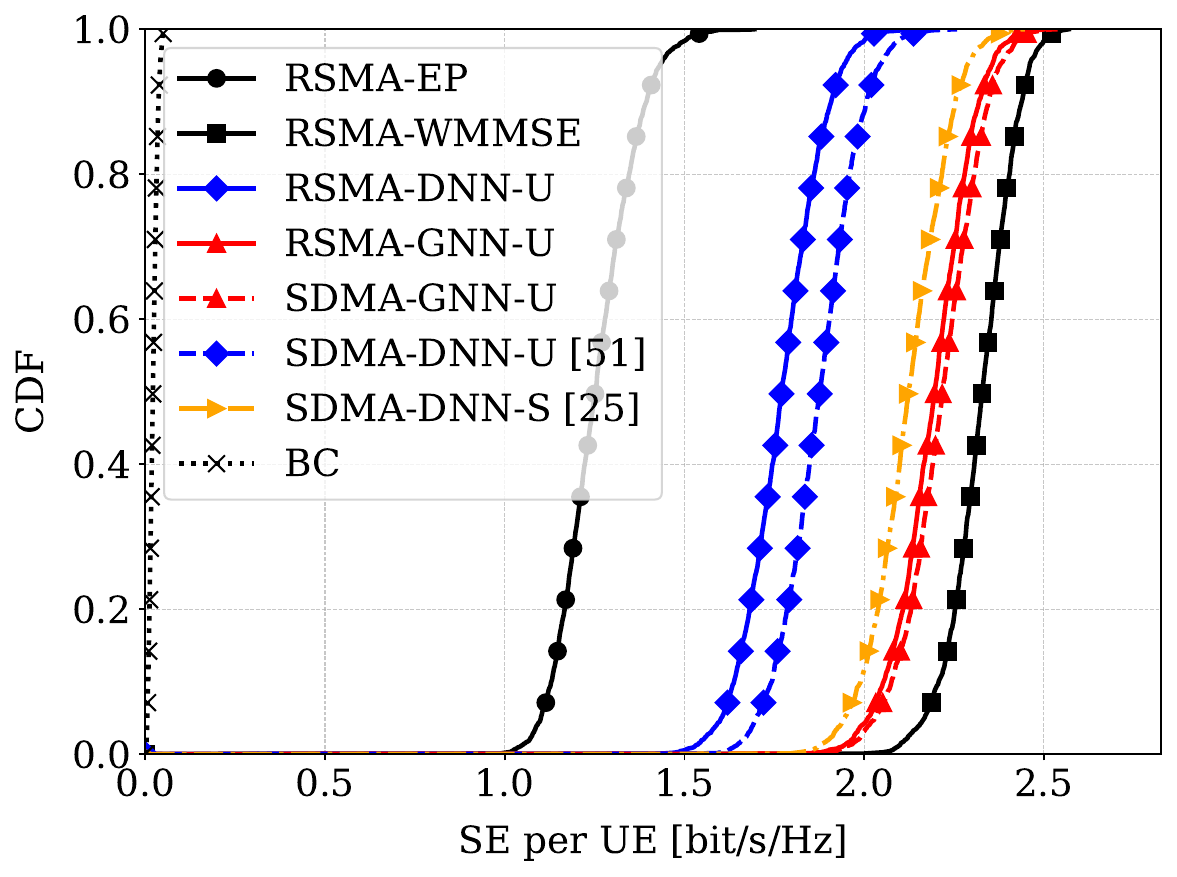}
    \par\vspace{0.5em}
    \footnotesize (c) Overloaded: $K=20$ UEs
    \label{fig:se_cdf_c}
\end{minipage}

\caption{CDF of the downlink SE per UE under different network loads: (a) Underloaded ($K=10$), (b) Fully Loaded ($K=16$), and (c) Overloaded ($K=20$) scenarios.
All plots are evaluated with $L=16$ APs and orthogonal pilot assignment.}
\label{fig:se_cdf_combined}
\end{figure*}

The CDF of the downlink SE per UE for the underloaded scenario is presented in Fig.~\ref{fig:se_cdf_combined}(a). As anticipated, the iterative \emph{RSMA-WMMSE} achieves the highest SE. In contrast, the \emph{BC} and \emph{RSMA-EP} schemes demonstrate the lowest performance due to their lack of spatial beamforming for effective interference mitigation. Regarding learning-based schemes, the supervised \emph{SDMA-DNN-S} exhibits a performance degradation of approximately 0.15 bit/s/Hz relative to \emph{RSMA-WMMSE}, due to the optimality gap inherent in the training labels. Notably, the proposed \emph{RSMA-GNN-U} and \emph{SDMA-GNN-U} closely approach the near-optimal \emph{RSMA-WMMSE}, maintaining a marginal gap of only 0.1 bit/s/Hz. Moreover, they outperform the unsupervised DNN counterparts by a significant margin of roughly 0.4 bit/s/Hz.

The evaluation extends to fully loaded and overloaded regimes in Fig.~\ref{fig:se_cdf_combined}(b) and (c). Despite intensified interference, the proposed framework remains robust. Notably, the SE gain of \emph{RSMA-GNN-U} over unsupervised DNN baselines widens from 0.4 bit/s/Hz in the fully loaded scenario to 0.5 bit/s/Hz in the overloaded case, highlighting the advantage of message-passing in capturing complex topologies. However, the specific gain of RSMA is marginal here, as the orthogonal pilot assignment inherently prevents pilot contamination. Regarding the optimality gap relative to the \emph{RSMA-WMMSE} benchmark, the proposed \emph{RSMA-GNN-U} exhibits remarkable stability, maintaining a narrow performance gap of only 0.15 bit/s/Hz.

\subsubsection{Convergence Analysis}
\begin{figure}[!t]
\centering
\includegraphics[width=3.1in]{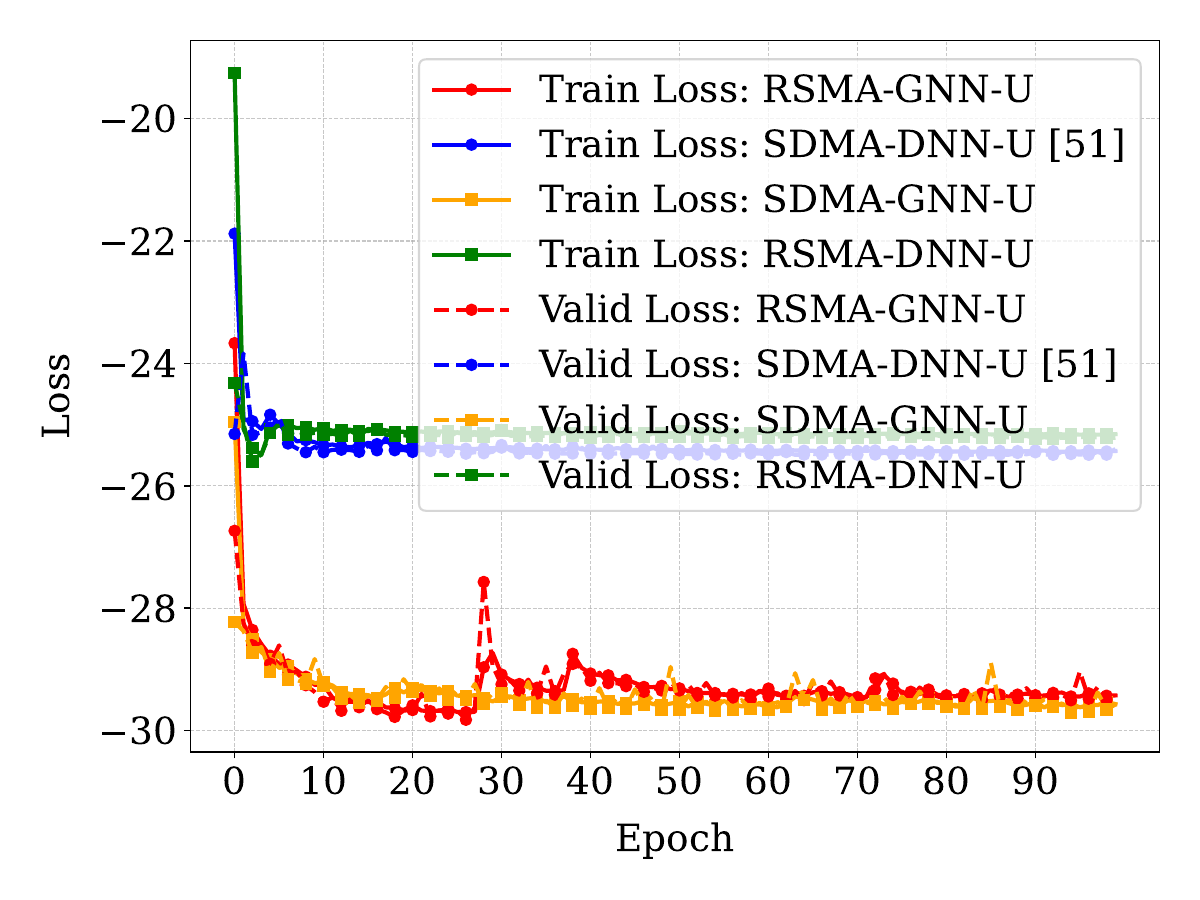}
\caption{Training and validation loss curves for the learning-based schemes over 100 epochs in the underloaded scenario ($L=16$ APs, $K=10$ UEs).}
\label{fig_un_loss3}
\end{figure}

Fig.~\ref{fig_un_loss3} illustrates the training and validation loss curves for the underloaded scenario, with similar trends observed across higher loads. Both losses decrease monotonically and converge within 40 epochs. Furthermore, the tight alignment between the training and validation curves confirms that the proposed framework effectively learns the channel characteristics without overfitting.

\subsubsection{Computational Complexity Analysis}
We evaluate the theoretical computational complexity in terms of floating-point operations (FLOPs). As established previously, the iterative \emph{RSMA-WMMSE} algorithm requires $\mathcal{O}\big(I_{\mathrm{out}} (K^3 L^3 + I_{\mathrm{in}} K^2 L^2)\big)$ operations, whereas the computational cost of the fully connected \emph{RSMA-DNN-U} scheme scales quadratically as $\mathcal{O}((LK)^2)$. In contrast, the proposed \emph{RSMA-GNN-U} achieves linear scalability. By leveraging a shared-parameter message-passing mechanism, its inference complexity across $T$ layers is bounded to $\mathcal{O}\left( T \cdot LK \cdot F_\mathrm{in}^2 \right)$. Furthermore, by integrating a user-centric clustering algorithm to sparsify the topology, this complexity can be further reduced to $\mathcal{O}\left( T \cdot \mathcal{E} \cdot F_\mathrm{in}^2 \right)$, where $\mathcal{E}$ is the number of active edges.

We empirically evaluate computational complexity via inference latency (averaged over 100 runs following 10 warm-ups) and trainable parameter count, summarized in Tables~\ref{tab_inference_time_transposed} and \ref{tab_para_transposed}. Consistent with \cite{2024arXiv_Benjamin}, the latency for \emph{SDMA-DNN-S} is measured on a per-AP basis.

\begin{table}[!t]
    \centering
    \caption{Inference Latency (\si{ms})}
    \label{tab_inference_time_transposed}
    \footnotesize
    \begin{tabular}{lccc}
        \hline
        \textbf{Scenarios (APs, UEs)} & \textbf{(16, 10)} & \textbf{(16, 16)} & \textbf{(16, 20)} \\
        \hline
        \textbf{RSMA-WMMSE}           & 2373.60          & 6329.39          & 7687.25          \\
        \textbf{RSMA-GNN-U}           & 5.63             & 7.21 & 9.69             \\
        \textbf{RSMA-DNN-U}           & \textbf{0.16}    & \textbf{0.14}    & \textbf{0.16}    \\
        \textbf{SDMA-GNN-U}           &5.62    & 7.18    & 9.77  \\
        \textbf{SDMA-DNN-U}           &\underline{0.18}  & \underline{0.17}    &\underline{0.18}  \\
        \textbf{SDMA-DNN-S}           &126.73           & 133.64           & 124.04           \\
        \hline
    \end{tabular}
\end{table}

\begin{table}[!t]
    \centering
    \caption{Trainable parameter count (K)}
    \label{tab_para_transposed}
    \footnotesize
    \begin{tabular}{lccc}
        \hline
        \textbf{Scenarios (APs, UEs)} & \textbf{(16, 10)} & \textbf{(16, 16)} & \textbf{(16, 20)} \\
        \hline
        \textbf{RSMA-WMMSE}           & --               & --               & --               \\
        \textbf{RSMA-GNN-U}      & 329.65     &594.70 & 821.37           \\
        \textbf{RSMA-DNN-U}           & 777.46           & 869.71            & 931.22           \\
        \textbf{SDMA-GNN-U}   & \underline{303.01}   &\underline{561.92}           & \underline{775.49}           \\
        \textbf{SDMA-DNN-U}           & 769.76           & 862.02           & 923.52           \\
        \textbf{SDMA-DNN-S}           & \textbf{79.04}   &\textbf{85.28}    & \textbf{89.44}            \\
        \hline
    \end{tabular}
\end{table}

As demonstrated in Table~\ref{tab_inference_time_transposed}, the proposed \emph{RSMA-GNN-U} reduces latency by three orders of magnitude compared to the iterative \emph{RSMA-WMMSE}. Specifically, the WMMSE solver requires nearly $7.7$~s to converge in the overloaded regime, while \emph{RSMA-GNN-U} executes in under $10$~ms. Although the fully-connected DNNs achieve sub-millisecond latencies ($0.16$~ms), the $5\text{-}10$~ms runtime of both GNN architectures easily satisfies typical $10\text{-}20$~ms channel coherence time constraints.

Regarding the trainable parameter count detailed in Table~\ref{tab_para_transposed}, \emph{RSMA-GNN-U} requires approximately 57\% fewer parameters than \emph{RSMA-DNN-U} in the underloaded setting, with \emph{SDMA-GNN-U} maintaining a similarly efficient footprint. While \emph{SDMA-DNN-S} uses the fewest parameters, it suffers from poor generalization and relies on computationally expensive label generation. Ultimately, the proposed \emph{RSMA-GNN-U} delivers near-optimal SE with real-time latency, while circumventing both the prohibitive execution time of iterative solvers and the parameter explosion of conventional DNNs.
  
\subsection{Performance Evaluation with Mixed-Scenario Training}
In this subsection, we assess the performance of the proposed framework under the \textit{Mixed-Scenario Training} regime, where a unified GNN model is trained to generalize across varying topologies via an adaptive embedding layer. To rigorously validate the scalability and robustness of the model, we conduct controlled experiments across four distinct dimensions: (i) scalability with respect to AP density, (ii) scalability with respect to user loading, (iii) adaptability to pilot configurations, and (iv) out-of-distribution (OOD) generalization, assessing performance in system configurations (e.g., $L, K$, and pilot lengths) diverge from the training distribution.

\subsubsection{Scalability with Respect to AP Density} 
\begin{figure}[!t]
\centering
\begin{minipage}{0.48\columnwidth}
    \centering
    \includegraphics[width=\linewidth]{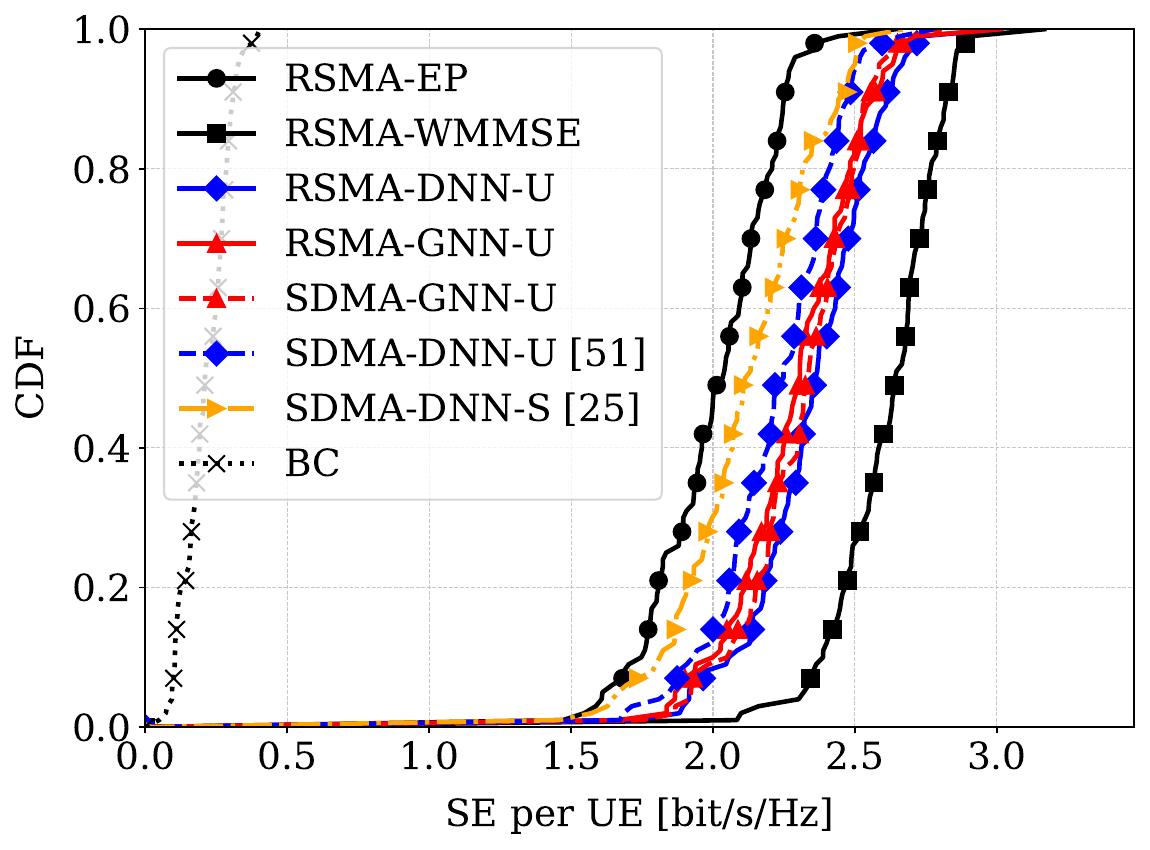}
    \footnotesize (a) $L=9$ APs
    \label{fig_en_se1}
\end{minipage}
\hfill
\begin{minipage}{0.48\columnwidth}
    \centering
    \includegraphics[width=\linewidth]{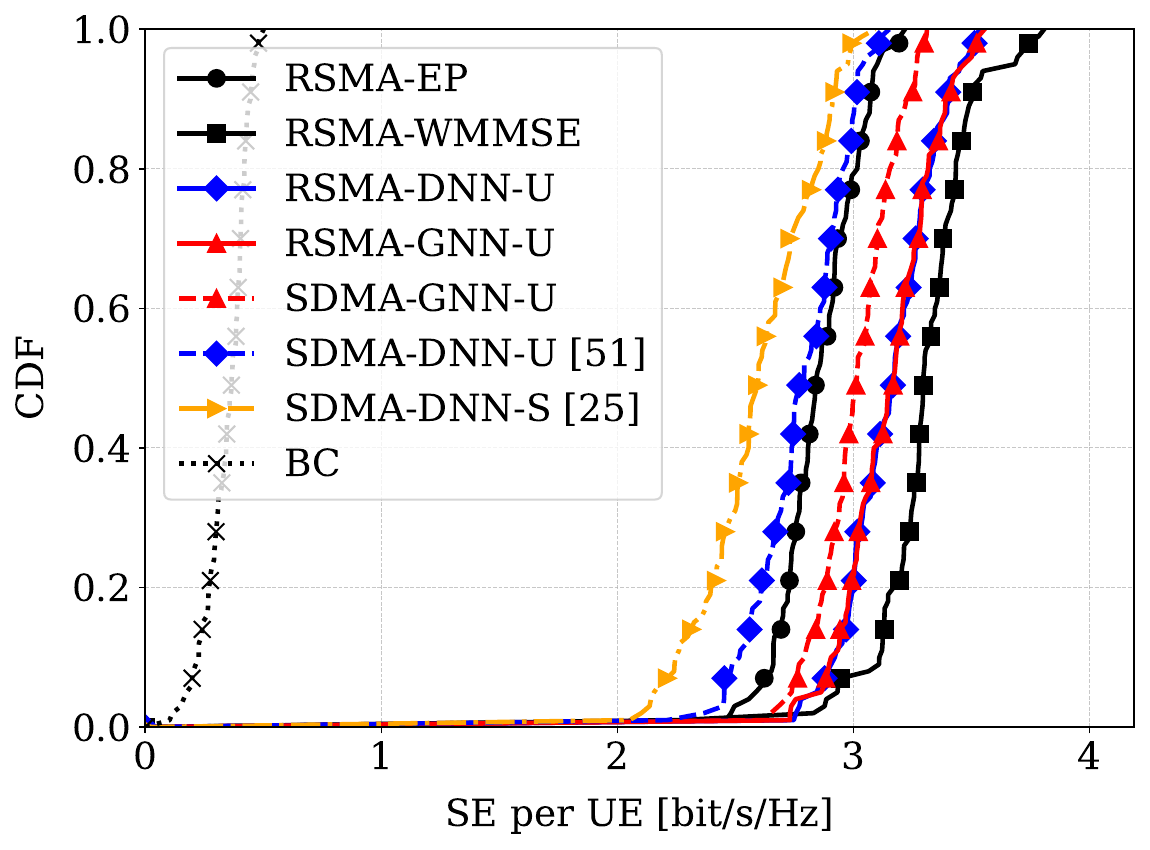}
    \footnotesize (b) $L=16$ APs
    \label{fig_en_se2}
\end{minipage}

\vspace{0.3cm} 

\begin{minipage}{0.48\columnwidth}
    \centering
    \includegraphics[width=\linewidth]{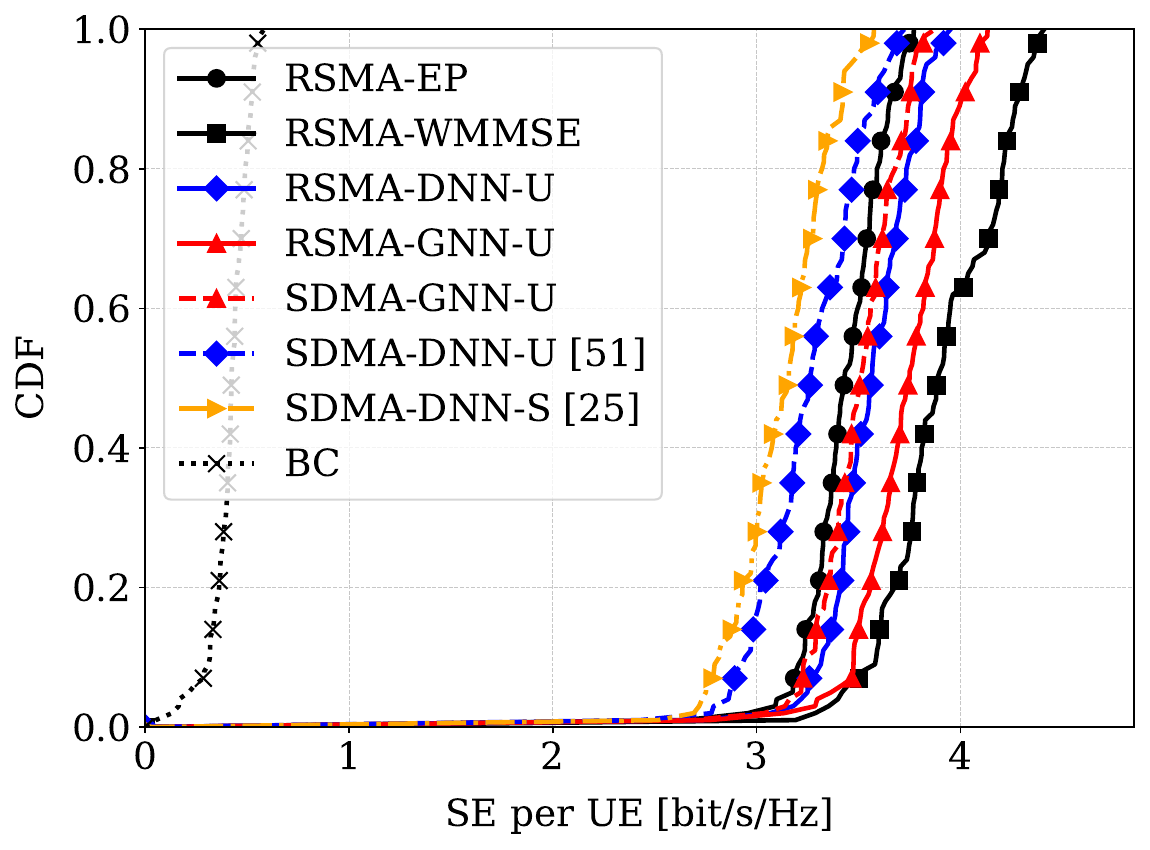}
    \footnotesize (c) $L=25$ APs
    \label{fig_en_se3}
\end{minipage}
\hfill
\begin{minipage}{0.48\columnwidth}
    \centering
    \includegraphics[width=\linewidth]{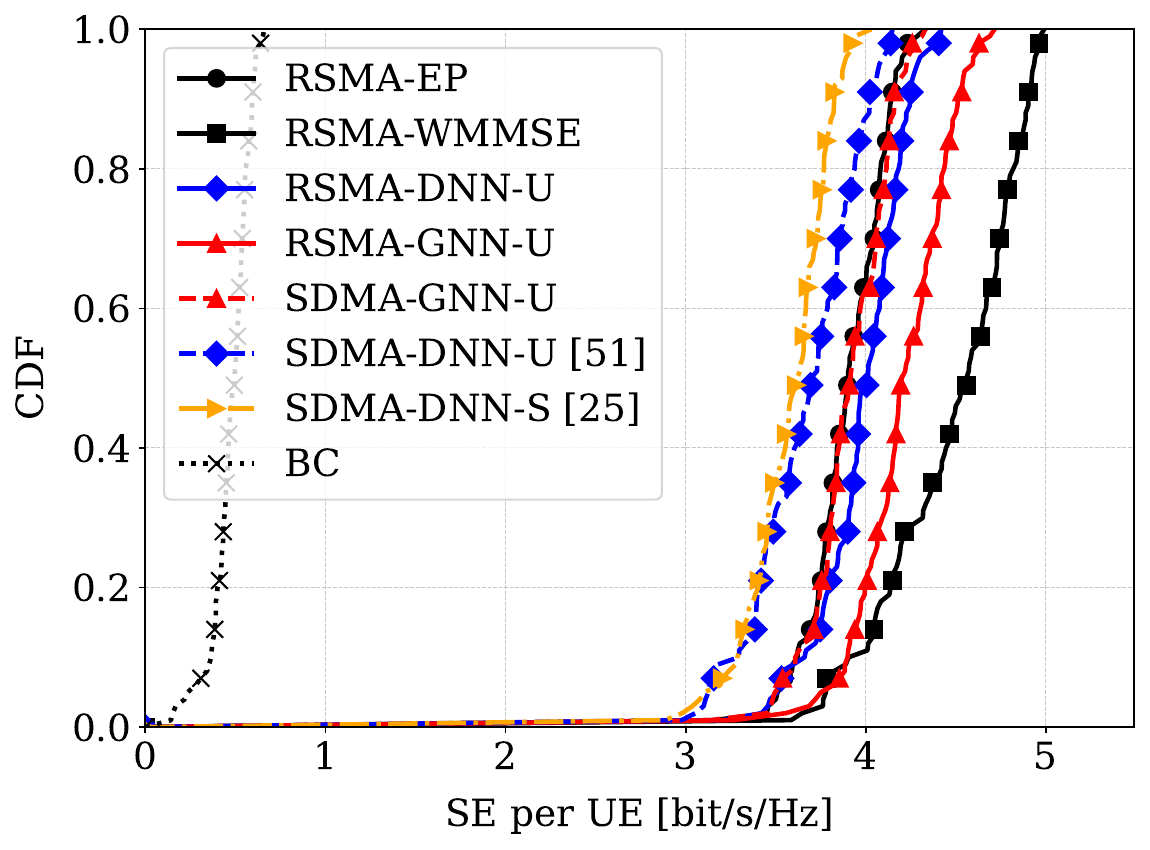}
    \footnotesize (d) $L=36$ APs
    \label{fig_en_se4}
\end{minipage}
\caption{CDF of the downlink SE per UE with $K = 6$ UEs and varying numbers of 
APs under orthogonal pilot assignment.}
\label{fig_en_se}
\end{figure}

We investigate the scalability of the unified model by inferencing on topologies with a fixed user load of $K=6$ and varying AP counts $L \in \{9, 16, 25, 36\}$, as illustrated in Fig.~\ref{fig_en_se}. 

As observed, both GNN-based architectures exhibit robust scalability as network density increases. Specifically, \emph{RSMA-GNN-U} closely tracks the near-optimal \emph{RSMA-WMMSE} benchmark with a stable gap of $0.1\text{-}0.3\,\text{bit/s/Hz}$, while consistently outperforming its counterpart (\emph{SDMA-GNN-U}), verifying the SE gains of rate-splitting. This stability contrasts sharply with the DNN baselines (\emph{RSMA-DNN-U} and \emph{SDMA-DNN-U}). While GNN and DNN schemes perform comparably in the sparse regime ($L=9$), the DNNs degrade significantly as the network scales. These results underscore both the superior topological generalization of the GNN framework and the inherent limitations of fixed-size DNN architectures in dense deployments.

\subsubsection{Scalability with Respect to User Loading} 
We next investigate the scalability with respect to UE density. The unified model is evaluated on topologies with fixed $L=16$ APs and varying user populations $K \in \{6, 10, 16, 20\}$. 
\begin{figure}[!t]
\centering
\begin{minipage}{0.48\columnwidth}
    \centering
    \includegraphics[width=\linewidth]{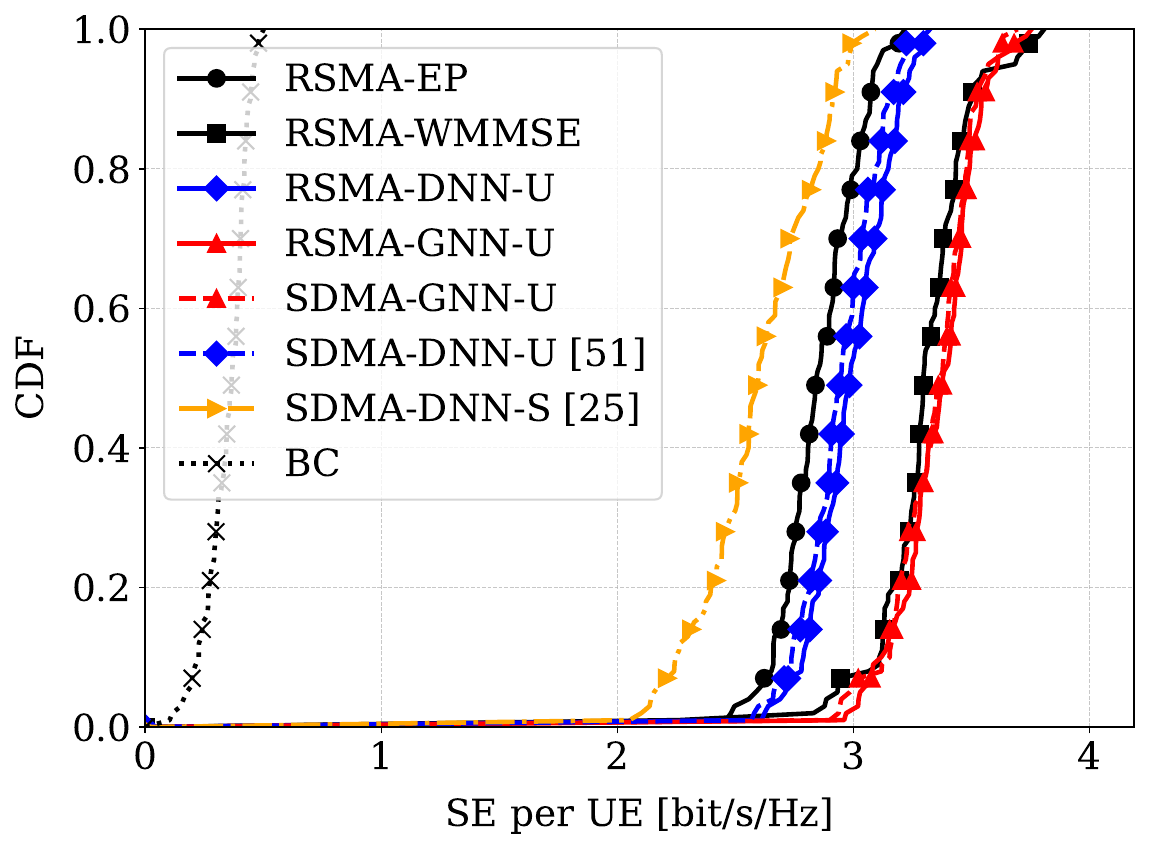}
    \footnotesize (a) $K=6$ UEs
    \label{fig_en_se5}
\end{minipage}
\hfill
\begin{minipage}{0.48\columnwidth}
    \centering
    \includegraphics[width=\linewidth]{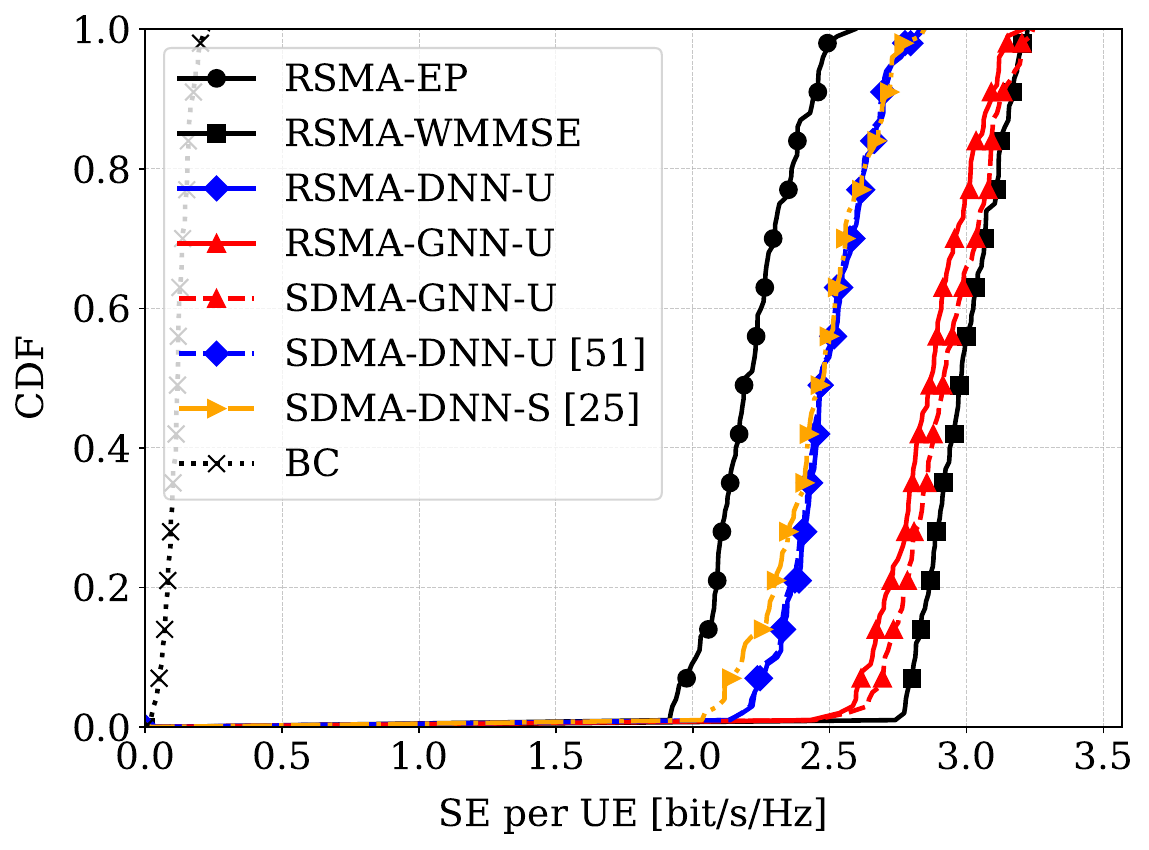}
    \footnotesize (b) $K=10$ UEs
    \label{fig_en_se6}
\end{minipage}

\vspace{0.3cm} 

\begin{minipage}{0.48\columnwidth}
    \centering
    \includegraphics[width=\linewidth]{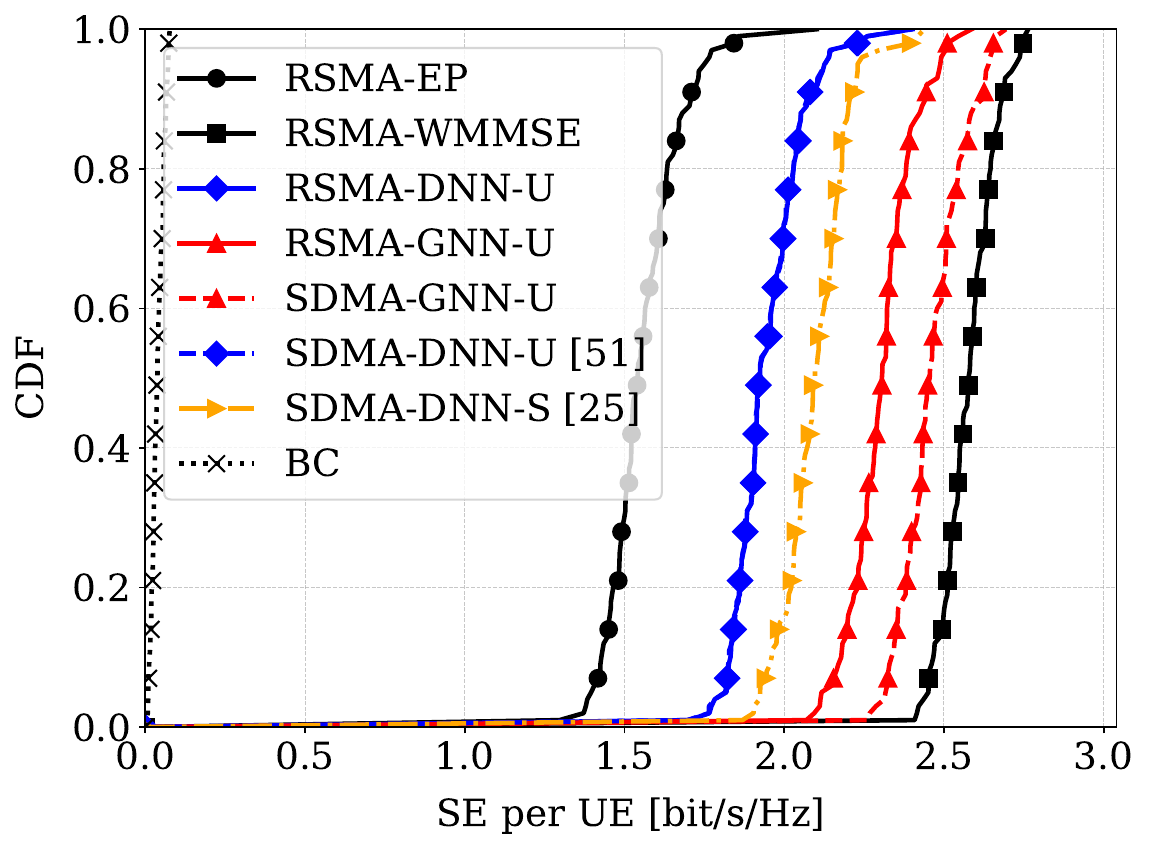}
    \footnotesize (c) $K=16$ UEs
    \label{fig_en_se7}
\end{minipage}
\hfill
\begin{minipage}{0.48\columnwidth}
    \centering
    \includegraphics[width=\linewidth]{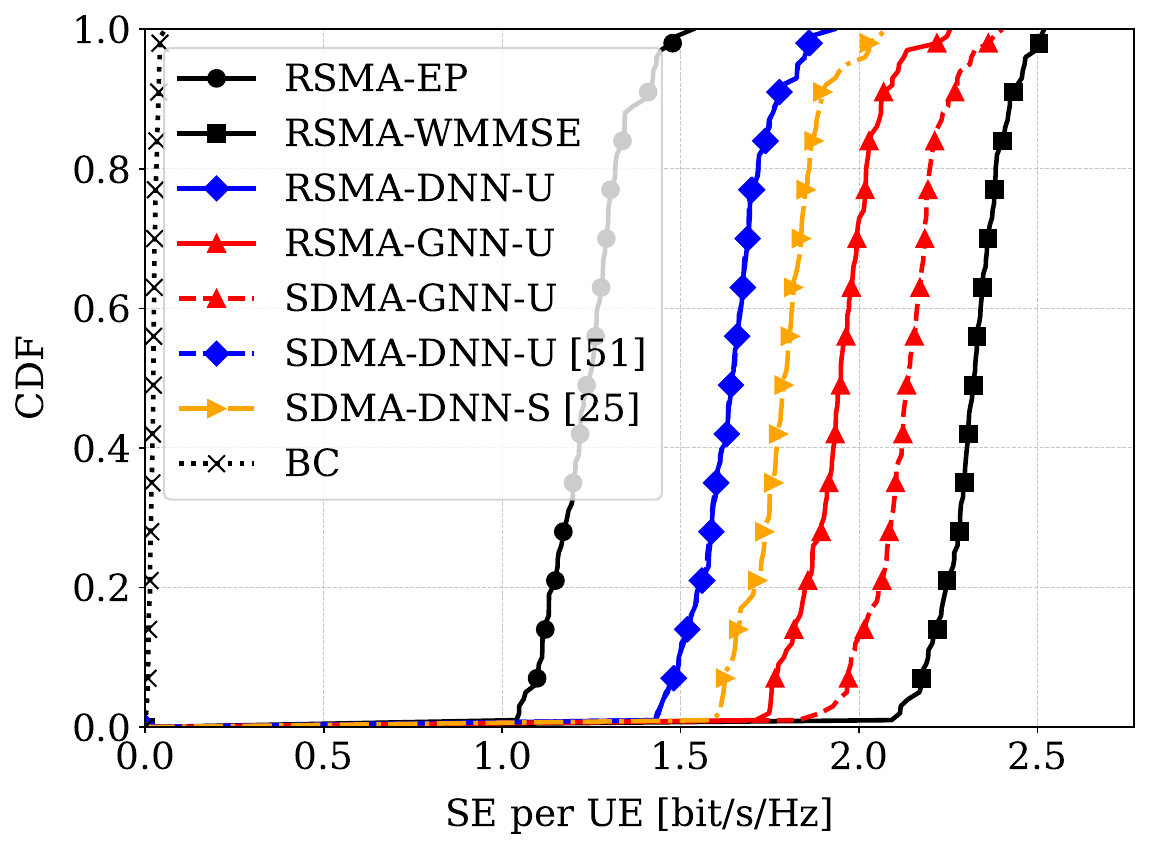}
    \footnotesize (d) $K=20$ UEs
    \label{fig_en_se8}
\end{minipage}

\caption{CDF of the downlink SE per UE with $L = 16$ APs and varying numbers of 
UEs under orthogonal pilot assignment.}
\label{fig_en_se_ue}
\end{figure}

As depicted in Fig.~\ref{fig_en_se_ue}, the GNN-based frameworks demonstrate remarkable resilience to interference escalation. Interestingly, under this orthogonal pilot assignment scheme, \emph{SDMA-GNN-U} slightly outperforms \emph{RSMA-GNN-U}, maintaining a tighter optimality gap relative to the \emph{RSMA-WMMSE} benchmark. This phenomenon reveals that when pilot contamination is inherently prevented, the powerful spatial interference mitigation capability of the GNN renders the common stream in RSMA redundant, making pure SDMA more power-efficient. 

In contrast, the fixed-size DNN architectures experience severe performance degradation as the number of UEs increases. For instance, the optimality gap for \emph{RSMA-DNN-U} widens drastically from $0.3\,\text{bit/s/Hz}$ to over $0.5\,\text{bit/s/Hz}$, underscoring the limited capacity of conventional DNNs to resolve high-dimensional multi-user interference. Ultimately, both adaptive GNN architectures maintain a commanding lead over the DNN baselines and heuristic approaches across all evaluated loads.

\subsubsection{Generalization to Different Pilot Assignment Schemes} 
\begin{figure*}[!t]
\centering
\begin{minipage}[b]{0.32\textwidth} 
    \centering
    \includegraphics[width=\textwidth]{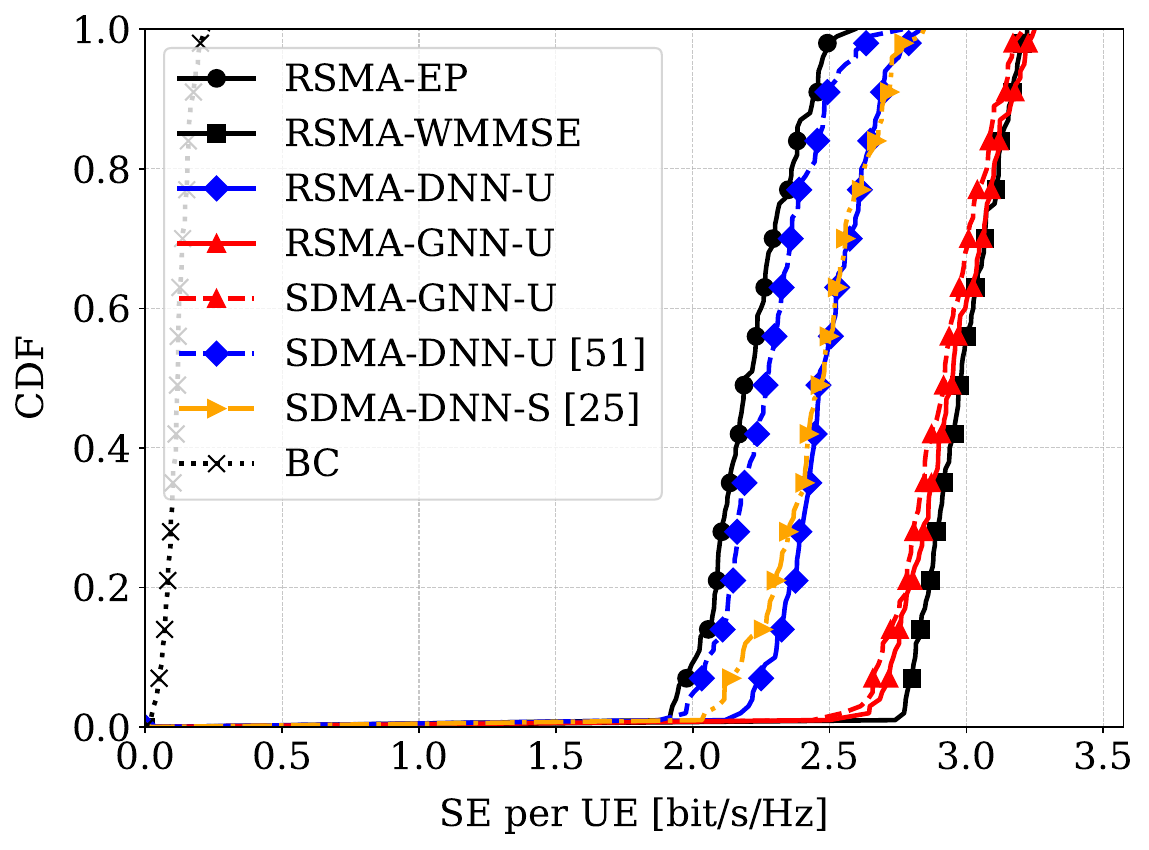}
    \par\vspace{0.5em}
    \footnotesize (a) Sufficient Pilot Resources: $K=10$ 
    \label{fig:p_se_cdf_a}
\end{minipage}
\hspace{-3mm} 
\begin{minipage}[b]{0.32\textwidth}
    \centering
    \includegraphics[width=\textwidth]{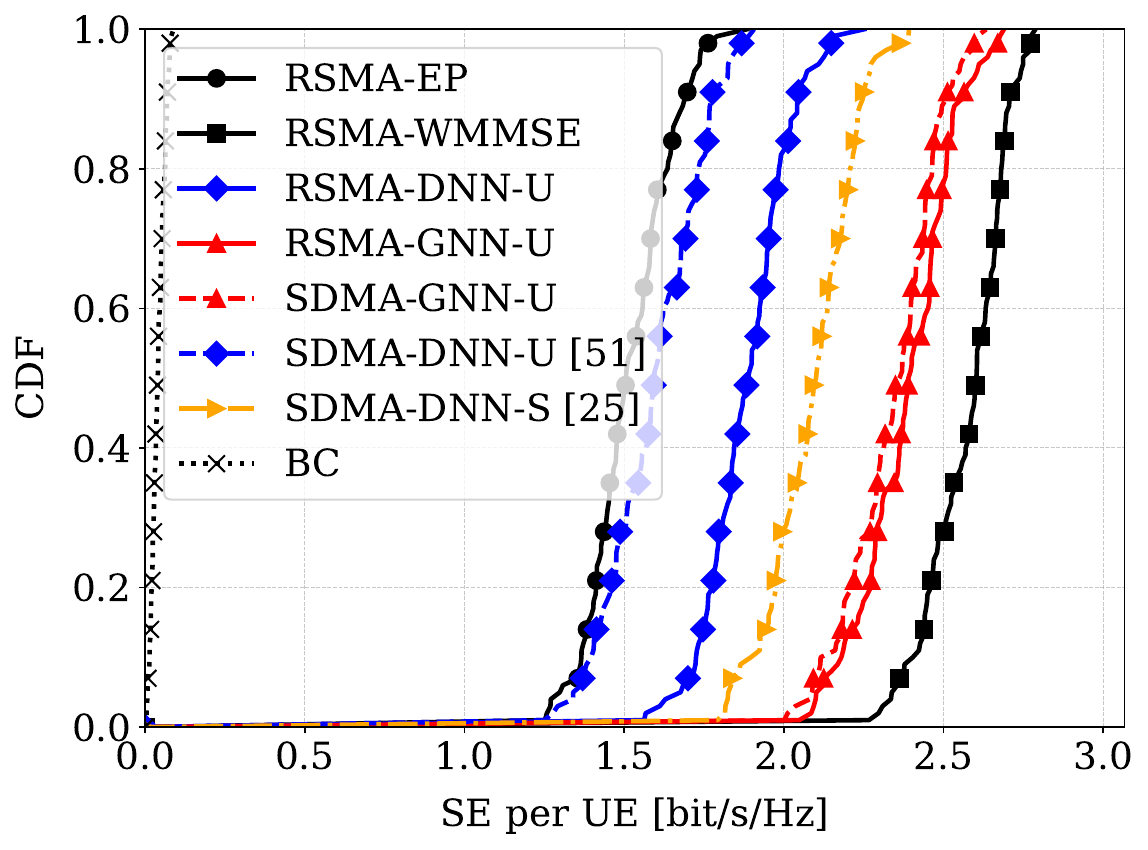}
    \par\vspace{0.5em}
    \footnotesize (b) Strained Pilot Resources: $K=16$
    \label{fig:p_se_cdf_b}
\end{minipage}
\hspace{-3mm} 
\begin{minipage}[b]{0.32\textwidth}
    \centering
    \includegraphics[width=\textwidth]{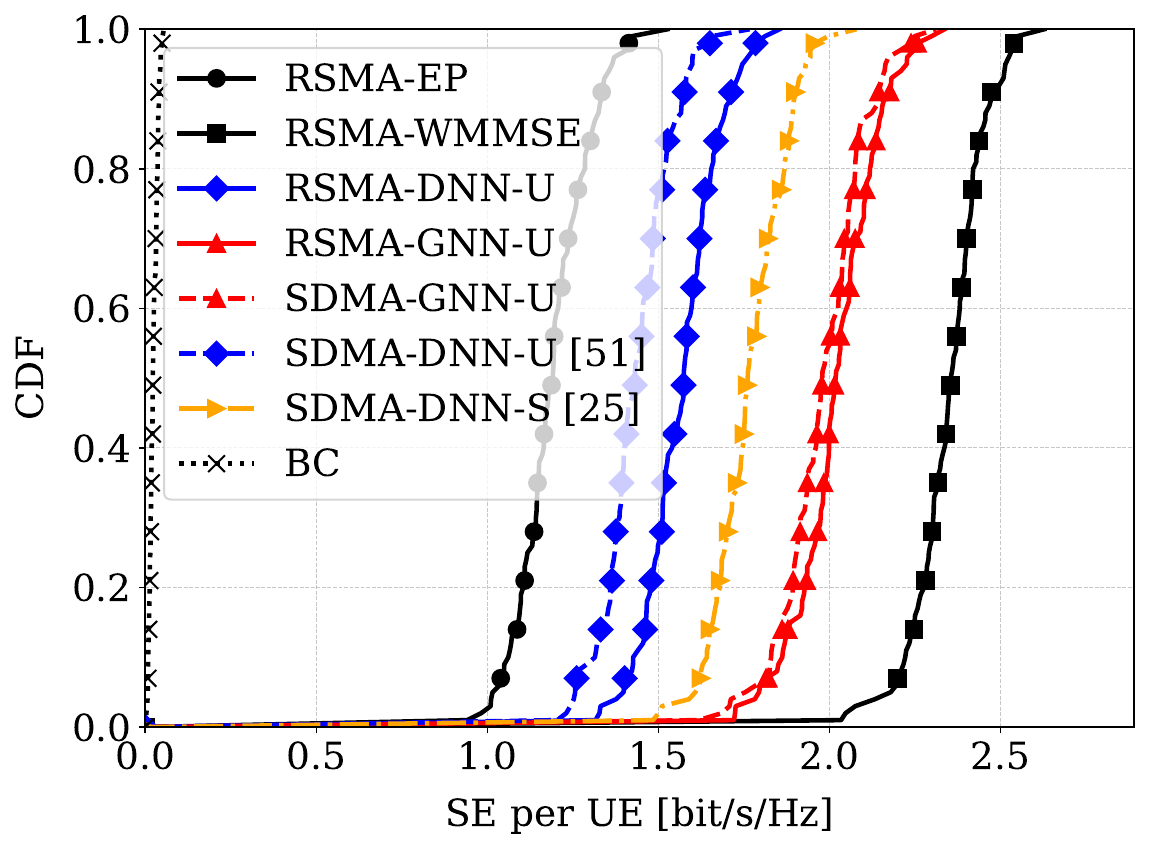}
    \par\vspace{0.5em}
    \footnotesize (c) Severely Strained Pilot Resources: $K=20$
    \label{fig:p_se_cdf_c}
\end{minipage}

\caption{CDF of the downlink SE per UE under varying Pilot Resource Strains with fixed $\tau_p = 10$: (a) Sufficient Pilot Resources ($K = 10$), (b) Strained Pilot Resources ($K = 16$), (c) Severely Strained Pilot Resources ($K=20$).
All plots are evaluated with $L=16$ APs.}
\label{fig_p_se}
\end{figure*}

We further assess the generalization capability of the scalable model under different pilot assignment schemes. In this experiment, the pilot length is fixed at $\tau_p = 10$, and the AP count at $L=16$, while the number of UEs varies across $K \in \{10, 16, 20\}$. Consequently, the system transitions from a contamination-free regime ($K \le \tau_p$) to a pilot-contaminated regime ($K > \tau_p$). The corresponding CDFs of the downlink SE per UE are depicted in Fig.~\ref{fig_p_se}.
% === 并排图片排版 ====

As illustrated in Fig.~\ref{fig_p_se}, the proposed \emph{RSMA-GNN-U} demonstrates exceptional resilience to pilot contamination and outperforms its counterpart \emph{SDMA-GNN-U}. Even as channel estimation errors become pronounced, the GNN model maintains a tight performance alignment with the \emph{RSMA-WMMSE} benchmark, keeping the SE optimality gap within $0.3\,\text{bit/s/Hz}$. This suggests that the GNN effectively learns the interference structure even with imperfect CSI. By contrast, the DNN-based schemes and \emph{RSMA-EP} experience significant performance deterioration in the non-orthogonal case, reflecting their limited ability in mitigating the impact of multi-user interference and pilot contamination.

\subsubsection{Out-of-Distribution Generalization to Unseen Configurations}
We finally assess the model’s scalability by evaluating its ability to generalize to unseen system configurations. As detailed in Table~\ref{tab:system_configurations}, the model is trained exclusively on medium-scale deployments but evaluated on unseen small- and large-scale scenarios with concurrently varying APs, UEs, and pilot assignments.

\begin{table}[t]
    \centering
    \caption{System Configurations for Model Training and Generalization Testing}
    \label{tab:system_configurations} % 建议使用这个新的标签
    \begin{tabular}{lcccc}
        \toprule % 顶部粗线
        \textbf{Scenario Type} & \textbf{Scenario ID} & \textbf{APs ($L$)} & \textbf{UEs ($K$)} & \textbf{Pilots ($\tau_p$)} \\
        \midrule % 中等粗细的线
        \multirow{8}{*}{\textbf{Training}} 
         & T1 & 16 & 6 & 6 \\
         & T2 & 16 & 10 & 10 \\
         & T3 & 16 & 16 & 10 \\
         & T4 & 16 & 16 & 16 \\
         & T5 & 25 & 6 & 6 \\
         & T6 & 25 & 10 & 10 \\
         & T7 & 25 & 16 & 10 \\
         & T8 & 25 & 16 & 16 \\
        \midrule % 分割训练集和测试集
        \multirow{5}{*}{\textbf{Testing}} 
         & E1 & 9 & 6 & 6 \\
         & E2 & 16 & 20 & 10 \\
         & E3 & 16 & 20 & 20 \\
         & E4 & 36 & 6 & 6 \\
         & E5 & 36 & 10 & 10 \\
        \bottomrule % 底部粗线
    \end{tabular}
    \footnotesize % 使用脚注来解释不变的参数
\end{table}

\begin{table*}[!t]
    \centering
    \caption{95\%-outage downlink SE per UE across different scenarios (bit/s/Hz)}
    \label{tab:general_scalability_table}
    \footnotesize
    \begin{tabular}{lcccccccc}
        \hline
        \textbf{Scenarios} & \textbf{RSMA-} & \textbf{RSMA-} & \textbf{RSMA-}  & \textbf{SDMA-} & {\textbf{SDMA-}} & {\textbf{SDMA-}} & {\textbf{RSMA-}} & {\textbf{BC}} \\
                                                {\textbf{(APs, UEs, Pilots)}}
                                                     & \textbf{WMMSE}& \textbf{GNN-U} & \textbf{DNN-U}  & \textbf{GNN-U}& \textbf{DNN-U} & \textbf{DNN-S} & \textbf{EP} & \\
        \hline
        (9, 6, 6) & \textbf{2.85} & 2.45 & \underline{2.52} & 2.45 & 2.50 & 2.50 & 2.28 & 0.33  \\
        (16, 20, 10) & \textbf{2.51} & 1.62 & 1.73 & 1.59  & 1.76 & \underline{1.93} & 1.36 & 0.04 \\
        (16, 20, 20) & \textbf{2.46} & 1.67 & 1.80  & 1.84 & 1.63 & \underline{1.97} & 1.43 & 0.04 \\
        (36, 6, 6) & \textbf{4.93} & 4.14 & \underline{4.27} & 4.04  & 4.20 & 3.88 & 4.20 & 0.61 \\
        (36, 10, 10) & \textbf{4.14} & 3.62 & 3.60 & 3.57 & \underline{3.69} & 3.22 & 3.55 & 0.29 \\
        \hline
    \end{tabular}
\end{table*}

The 95\%-outage downlink SE results, summarized in Table~\ref{tab:general_scalability_table}, validate the inherent robustness of the proposed framework. Despite operating entirely outside its training distribution, the unified \emph{RSMA-GNN-U} model exhibits highly stable edge-user performance. Notably, it closely trails the dimension-specific \emph{RSMA-DNN-U} benchmark with a maximum performance gap of only $0.13\,\text{bit/s/Hz}$, while offering massive computational savings. Moreover, comparing the multi-access mechanisms reveals that \emph{RSMA-GNN-U} generally matches or outperforms its SDMA counterpart (\emph{SDMA-GNN-U}) across most OOD topologies. This reaffirms that the rate-splitting architecture successfully preserves its robust interference mitigation capabilities even when generalized to entirely new network scales.

\subsubsection{Computational Complexity Analysis}
\begin{figure}[!t]
\centering
\includegraphics[width=3.1in]{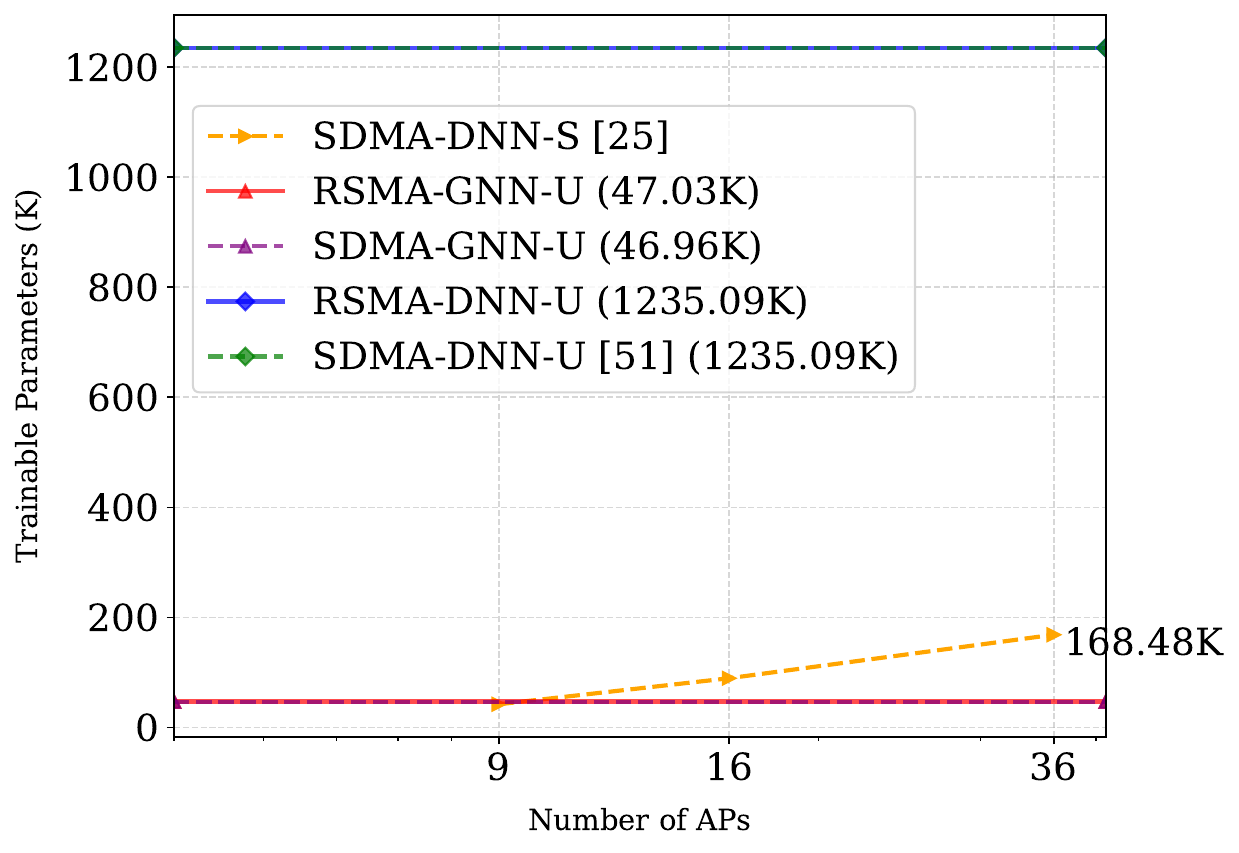}
\caption{Trainable parameter count across varying number of APs.}
\label{fig_en_para}
\end{figure}
Having established the theoretical complexity bounds previously, we now evaluate the empirical computational cost of the unified model across OOD topologies. As illustrated in Fig.~\ref{fig_en_para}, the trainable parameter space of the proposed GNN is entirely decoupled from the varying system dimensions, requiring zero architectural modifications for unseen scales.

\begin{table*}[!t]
    \centering
    \caption{Inference Latency across Different Scenarios (\si{ms})}
    \label{tab_inference_time_scala}
    \footnotesize
    \begin{tabular}{lccccccc}
        \hline
        \textbf{Scenarios} & \textbf{RSMA-} & \textbf{RSMA-} & \textbf{RSMA-}  & \textbf{SDMA-} & {\textbf{SDMA-}} & {\textbf{SDMA-}} &  \\
                                                {\textbf{(APs, UEs, Pilots)}}
                                                     & \textbf{WMMSE}& \textbf{GNN-U} & \textbf{DNN-U}& \textbf{GNN-U} & \textbf{DNN-U} & \textbf{DNN-S} &  \\
        \hline
        (9, 6, 6) & 14871 & 7.21 & \textbf{0.29} & 7.23 & \underline{0.31} & 388.62 \\
        (16, 20, 10) & 366439 & 37.48 & \textbf{0.24} & 37.52 & \underline{0.28} & 394.88 \\
        (16, 20, 20) & 371267 & 41.54 & \textbf{0.26} & 41.75 & \underline{0.29} & 390.73 \\
        (36, 6, 6) & 180176 & 37.76 & \textbf{0.25} & 31.73 & \underline{0.29} & 369.42 \\
        (36, 10, 10) & 59707 & 59.00 & \textbf{0.25} & 49.10 & \underline{0.28} & 374.43 \\
        \hline
    \end{tabular}
\end{table*}
As detailed in Table~\ref{tab_inference_time_scala}, both adaptive GNN architectures maintain remarkably low and predictable inference latencies, scaling smoothly from $7.2\,\text{ms}$ in sparse configurations to $59.0\,\text{ms}$ in massive topologies. Crucially, compared to the iterative \emph{RSMA-WMMSE} algorithm, which requires anywhere from $14.8\,\text{s}$ to over $371\,\text{s}$ to converge, the proposed \emph{RSMA-GNN-U} achieves a massive latency reduction of up to three orders of magnitude. Furthermore, the GNN framework accelerates inference by a factor of 6 to 50 relative to the supervised \emph{SDMA-DNN-S} baseline. These results confirm that the unified GNN natively satisfies the stringent real-time constraints of dynamic CF-mMIMO networks without sacrificing SE.

\subsection{Computational Efficiency of Scalable Architectures}
 To evaluate the computational efficiency and parameter scalability of our proposed slice-based mechanism, we benchmark it against two established baselines: an attention-weighted GNN and a dynamic pooling GNN. Table VIII summarizes the comparative performance in terms of trainable parameters and multiply-accumulate (MAC) operations.

\begin{table*}[!t]
    \centering
    \caption{Quantitative Comparison of Parameters and Computational Complexity (MACs) Across Different Scalable Architectures}
    \label{tab:architecture_complexity}
    \footnotesize
    \begin{tabular}{lcccccccc}
        \toprule
        \textbf{Scenarios (APs, UEs)} & \textbf{Global $D_{\mathrm{in}}$} & \textbf{Slice Params} & \textbf{Att. Params} & \textbf{Pool. Params} & \textbf{Slice MACs} & \textbf{Att. MACs} & \textbf{Pool. MACs} \\
        \midrule
        (16, 6)  & 22  & \textbf{47.03K} & 43.91K & 46.47K & \textbf{939.20K} & 1.30M & 3.75M \\
        (16, 10) & 26  & \textbf{47.03K} & 45.20K & 46.99K & \textbf{1.13M}   & 1.89M & 3.77M \\
        (25, 16) & 41  & \textbf{47.03K} & 49.42K & 48.34K & \textbf{1.88M}   & 4.97M & 3.85M \\
        (36, 32) & 68  & \textbf{47.03K} & 57.38K & 51.10K & \textbf{3.41M}   & 18.40M & 4.02M \\
        (64, 40) & 104 & \textbf{47.03K} & 67.11K & 53.93K & \textbf{5.84M}   & 56.85M & 4.31M \\
        (128, 80) & 208 & \textbf{47.03K} & 96.34K & 63.19K & \textbf{15.21M}  & 424.08M & 5.60M \\
        \bottomrule
    \end{tabular}
\end{table*}

As indicated in Table VIII, the attention baseline suffers from severe parameter inflation (from 43.91K to 96.34K) and quadratic complexity growth, with its computational burden exploding from 1.30M MACs to 424.08M MACs as the network scales. This is due to the direct scale dependence of its high-dimensional node attributes in expanding topologies. Meanwhile, the pooling baseline incurs high preprocessing sorting overhead and structural distortions during graph downsampling, with its parameter count growing from 46.47K to 63.19K. In contrast, the proposed slice-based method guarantees an absolute invariant model capacity of 47.03K parameters across all scaled deployments satisfying $D_{\mathrm{in}} \le D_{\max}$, while maintaining a highly competitive and strictly linear computational growth profile (scaling conservatively from 939.20K to 15.21M MACs). For extreme scaling scenarios where the network dimension exceeds the predefined operational boundary ($D_{\mathrm{in}} > D_{\max}$), the framework operates seamlessly alongside the user-centric clustering strategy introduced in Section IV-C. By decomposing the ultra-large continuous topology into size-bounded overlapping subgraphs ($D_{\mathrm{sub}} \le D_{\max}$), our neural architecture scales to arbitrary networks without retraining, maintaining full graph topological integrity while ensuring strict local parameter and execution efficiency.

\subsection{Feasibility Analysis in Ultra-Dense Scenarios}\label{sec:ultra_dense}
    
We evaluate the feasibility of the proposed framework in ultra-dense regimes. If the network is modeled as a fully connected bipartite graph, the message-passing complexity creates a fundamental computational bottleneck. However, severe path loss dictates that each UE is effectively served by only a sparse subset of APs. 

To evaluate the framework's adaptability to such extreme scales, we implement a user-centric graph sparsification strategy prior to GNN inference. By retaining only the $Q_\mathrm{UE}$ dominant large-scale links per UE, the message-passing complexity is strictly bounded to $\mathcal{O}(T \cdot |\mathcal{E}| \cdot F^2) = \mathcal{O}(T \cdot Q_\mathrm{UE} K \cdot F^2)$. 

\begin{table*}[t]
    \centering
    \caption{Scalability and Complexity Reduction Analysis (Configuration A: Max $D_{\text{sub}} = 64, Q_\mathrm{UE}=4$)}
    \label{tab:scalability_analysis_A_manuscript}
    \footnotesize
    \begin{tabular}{lcccccccc}
        \toprule
        \textbf{AP ($L$)} & \textbf{UE ($K$)} & \textbf{Dense $LK$} & \textbf{Active $|\mathcal{E}|$} & \textbf{$|\mathcal{E}|/LK$} & \textbf{Avg. AP/UE} & \textbf{Reduction ($LK/|\mathcal{E}|$)} & \textbf{Max $D_{\text{sub}}$} & \textbf{$T$-hop Pres.} \\
        \midrule
        128  & 128 & 16,384  & 512   & 3.12\% & 4.00 & 32.0$\times$  & 64 & 100\% \\
        256  & 160 & 40,960  & 640   & 1.56\% & 4.00 & 64.0$\times$  & 64 & 100\% \\
        512  & 320 & 163,840 & 1,280  & 0.78\% & 4.00 & 128.0$\times$ & 64 & 100\% \\
        1024 & 640 & 655,360 & 2,560  & 0.39\% & 4.00 & \textbf{256.0$\times$} & \textbf{64} & \textbf{100\%} \\
        \bottomrule
    \end{tabular}
\end{table*}

\begin{table*}[t]
    \centering
    \caption{Scalability and Complexity Reduction Analysis (Configuration B: Max $D_{\text{sub}} = 128, Q_\mathrm{UE}=8$)}
    \label{tab:scalability_analysis_B_manuscript}
    \footnotesize
    \begin{tabular}{lcccccccc}
        \toprule
        \textbf{AP ($L$)} & \textbf{UE ($K$)} & \textbf{Dense $LK$} & \textbf{Active $|\mathcal{E}|$} & \textbf{$|\mathcal{E}|/LK$} & \textbf{Avg. AP/UE} & \textbf{Reduction ($LK/|\mathcal{E}|$)} & \textbf{Max $D_{\text{sub}}$} & \textbf{$T$-hop Pres.} \\
        \midrule
        128  & 128 & 16,384  & 1,024 & 6.25\% & 8.00 & 16.0$\times$  & 128 & 100\% \\
        256  & 160 & 40,960  & 1,280 & 3.12\% & 8.00 & 32.0$\times$  & 128 & 100\% \\
        512  & 320 & 163,840 & 2,560 & 1.56\% & 8.00 & 64.0$\times$  & 128 & 100\% \\
        1024 & 640 & 655,360 & 5,120 & 0.78\% & 8.00 & \textbf{128.0$\times$} & \textbf{128} & \textbf{100\%} \\
        \bottomrule
    \end{tabular}
\end{table*}

Tables \ref{tab:scalability_analysis_A_manuscript} and \ref{tab:scalability_analysis_B_manuscript} present the boundary analysis under two distinct clustering configurations, scaling up to an extreme deployment of $L=1024$ APs and $K=640$ UEs. Under the first configuration (Table \ref{tab:scalability_analysis_A_manuscript}), the active edges constitute merely $0.39\%$ of the dense formulation, yielding a staggering $256\times$ reduction in computational cost. Crucially, local clustering restricts the peak subgraph dimension ($D_{\text{sub}} \le 64$), securing $100\%$ preservation of the $T$-hop receptive field. When a denser local association is demanded ($Q_\mathrm{UE}=8$, Table \ref{tab:scalability_analysis_B_manuscript}), scaling the offline capacity to $D_{\max}=128$ restores structural feasibility, achieving a $128\times$ complexity reduction while preserving perfect $T$-hop context. These results confirm that the scalability boundary is strictly governed by the local density $Q_\mathrm{UE}$ and offline capacity $D_{\max}$, thereby completely circumventing the global dimensionality curse.

\section{Conclusion}
In this paper, we proposed an unsupervised and scalable GNN framework for downlink power allocation in RS-CF-mMIMO systems, aiming to maximize system SE under different pilot assignment schemes and varying network scales. By modeling APs and UEs as graph nodes with large-scale coefficients as features, the framework preserved permutation equivariance and maintained low complexity in terms of both inference latency and trainable parameter count. A WMMSE-ADMM power allocation algorithm was developed that served as a performance upper bound. A key innovation of our work lay in the design of adaptive embedding layers, which enabled the scalable GNN architecture to generalize effectively across diverse network sizes and pilot configurations without the need for retraining. Numerical results demonstrated that the proposed scheme achieved near-optimal spectral efficiency and consistently outperformed unsupervised DNN-based models, supervised learning baselines, and heuristic power allocation strategies across underloaded, fully loaded, and overloaded scenarios. Finally, the comprehensive complexity analysis confirmed its practicality for real-time edge deployment.

\bibliographystyle{IEEEtran}   
\def\IEEEbibitemsep{0pt}

\bibliography{IEEEabrv,main}  
\end{document}